\documentclass[12pt,a4paper]{article}
\pdfoutput=1

\usepackage{jheppub}
\usepackage{color}
\usepackage{tikz}
\usepackage{bm}
\usepackage{amsmath}
\usepackage{amssymb}
\usepackage{amsthm}
\usepackage{ccaption}
\usepackage{subcaption}
\usepackage{graphicx}
\usepackage[super]{nth}
\usepackage{microtype}
\usepackage{cleveref}
\usepackage{listings}
\usepackage{dsfont}

\newcommand{\ket}[1]{|#1\rangle}
\newcommand{\braket}[2]{\langle #1|#2\rangle}

\graphicspath{{Figures/}}

  \captionnamefont{\bfseries}
  \captiontitlefont{\small\sffamily}
  \captiondelim{: }
  \hangcaption

\author{Alex Peach and Simon F. Ross}
\affiliation{Centre for Particle Theory \& Department of Mathematical Sciences,\\
                     Durham University, South Road, Durham DH1 3LE, UK.}
                     
\abstract{We study the entanglement structure of states dual to multiboundary wormhole geometries using tensor network models. Perfect and random tensor networks tiling the hyperbolic plane have been shown to provide good models of the entanglement structure in holography. We extend this by quotienting the plane by discrete isometries to obtain models of the multiboundary states. We show that there are networks where the entanglement structure is purely bipartite, extending results obtained in the large temperature limit. We analyse the entanglement structure in a range of examples.}

\title{Tensor Network Models of Multiboundary Wormholes}

\begin{document}
\maketitle

\section{Introduction}

Multiboundary wormhole geometries are a useful laboratory for studying the relation between the entanglement structure of CFT states and the bulk geometry. In \cite{Balasubramanian:2014hda}, an investigation of the entanglement structure of a class of asymptotically AdS $2+1$ dimensional spacetimes with $n$ asymptotic boundaries was initiated; these are dual to states in $n$ copies of the CFT on $S^1 \times \mathbb R$. These solutions were introduced in \cite{Brill:1995jv,Aminneborg:1997pz,Brill:1998pr,Aminneborg:1998si}, and their holographic study was initiated by \cite{Krasnov:2000zq,Krasnov:2003ye,Skenderis:2009ju}. The CFT states are given by a path integral on a Riemann surface $\Sigma$ with $n$ boundaries. The entanglement structure of these states has a complicated dependence on the moduli of the Riemann surface, exhibiting regions of multipartite entanglement but also regions where bipartite entanglement between different copies is dominant. In \cite{Marolf:2015vma}, the entanglement structure in a region of large moduli, where the CFT states involve highly excited states on each $S^1 \times \mathbb R$ factor, was explored in more detail. This is effectively a regime of high temperature, although the reduced density matrix in a single copy of the CFT is not necessarily thermal. The structure in this regime is dominated by local, bipartite entanglement between subregions on each boundary on a scale set by the effective temperature. There can be a multipartite component in this regime, but it is associated just with a single thermal volume, so it is a small part of the overall state. 

It is difficult to gain more insight into the entanglement structure for more generic moduli from the full CFT path integral. This motivates the study of tensor network models, which share many of the entanglement and geometrical features of the full state.\footnote{The most robust model of the holographic entanglement structure are MERA networks \cite{2009arXiv0912.1651V,EV1}, which provide a good description of the ground state in conformal field theories, and have been related to holographic descriptions of the state \cite{Swingle:2009bg,Swingle:2012wq,Czech:2015qta,Czech:2015kbp}. A MERA version of the quotient giving BTZ was constructed in \cite{Czech:2015xna}. However, it appears difficult to extend this construction to the more general quotients we are interested in with multiple generators. We will therefore focus our attention on more phenomenological models.} Other approaches to multiboundary wormholes have recently been explored in \cite{Salton:2016qpp,Balasubramanian:2016sro}; see also the interesting work on multipartite entanglement in tensor networks \cite{Nezami:2016zni}. 

The models we consider were introduced in \cite{Pastawski:2015qua,Yang:2015uoa,Hayden:2016cfa} to model the relation of the entanglement structure of the vacuum state to global AdS, explicitly exhibiting the ideas of code subspaces in \cite{Almheiri:2014lwa}. They are based on tiling the hyperbolic plane with perfect or random tensors, and were shown to reproduce the Ryu-Takayanagi formula for entanglement entropies. Following \cite{Bhattacharyya:2016hbx}, we consider discrete quotients of the tiled plane, and use the tensor network on the quotient space as a model of the CFT states dual to such multiboundary geometries. In these models, we can explore intermediate regions in the moduli space of Riemann surfaces, and study the entanglement structure of the corresponding states. 

Surprisingly, we find that even at generic values of the moduli, there can be tilings where the entanglement structure is purely bipartite. Although this result presumably reflects the limited resolution of the discrete tensor network models, it is interesting as it provides explicit illustration of the way in which a connected multiboundary state can be built up from purely bipartite entanglement. In these cases, the state is distillable to a state containing just Bell pairs. For other tilings, there is a residual multipartite component, and we attempt to characterise the multipartite structure in its entanglement using negativity of the reduced density matrices, comparing to a random state on the same Hilbert space. Further characterisation of this multipartite component is an interesting challenge for further work. Our computations are for low bond dimension, and it would also be interesting to see how they extend to higher bond dimension. 

In the next section, we review previous work on multiboundary wormholes. In section \ref{tiling}, we construct tilings of the Riemann surface $\Sigma$ for discrete values of the moduli by quotienting tilings of the hyperbolic plane by discrete isometries. We discuss the discrete analogue of horizons and the causal shadow region in these tilings, and show that in some cases there is no causal shadow region. In section \ref{networks}, we review the tensor network models built on the tilings of the hyperbolic plane. In section \ref{multi}, we apply these methods to the tilings of the Riemann surface $\Sigma$ and analyse the entanglement structure of the resulting states.    

\section{Holographic Multiboundary Wormholes}
\label{review}

The holographic description of multiboundary wormholes generalises the relation between the thermofield double state 
\begin{equation}
\label{TFD}
\ket{TFD}= \sum_E e^{-\frac{\beta}{2}E }\  \ket{E}_1 \ \ket{E}_2
\end{equation}
in two copies of the CFT and the eternal black hole \cite{Maldacena:2001kr}. This state is obtained as the result of a Euclidean CFT path integral on a cylinder of length $\beta/2$ (taking the $S^1$ to have period $2\pi$). The trace over one copy gives a thermal density matrix, at inverse temperature $\beta$. At sufficiently high temperatures (small $\beta$), the dominant bulk saddle for these boundary conditions is a Euclidean black hole. Analytically continuing to Lorentzian time, the two copies of the CFT live on the two boundaries of the black hole, and the entanglement of the state \eqref{TFD} is essential to account for the connectedness of the bulk geometry. 

In \cite{Balasubramanian:2014hda}, this picture was extended to consider the role of the entanglement in the CFT in multiboundary wormhole geometries. In $2 +1$ dimensions, such geometries can easily be constructed by considering quotients of vacuum AdS$_3$. The Euclidean quotients we are interested in are usefully described by writing the Euclidean $AdS_3$, equivalently $H^3$, in a coordinate system
\begin{equation}
\label{H3}
\frac{ds^2}{l_{AdS}^2} = dt_E^2 + \cosh^2 t_E d\Sigma^2
\end{equation}
where $t_E$ is Euclidean time and $d \Sigma^2$ is the unit-radius metric on $H^2$. The eternal BTZ black hole arises as a quotient by a discrete subgroup $\Gamma$ of the $SL(2,\mathbb R)$ isometry group  of this $H^2$ generated by a single hyperbolic element \cite{Banados:1992gq}. This converts $H^2$ into a cylinder with two boundaries, with a hyperbolic metric. The more general quotients we are interested in correspond to considering discrete subgroups $\Gamma$ generated by $k$ hyperbolic elements. These geometries were introduced in \cite{Brill:1995jv,Aminneborg:1997pz,Brill:1998pr,Aminneborg:1998si}. The resulting surface $\Sigma = H^2/\Gamma$ is a smooth Riemann surface with genus $g$ and $n$ boundaries. This Riemann surface has $6g - 6 + 3n$ moduli, which are encoded in the choice of discrete group $\Gamma$. Since the quotient acts on the surfaces of constant $t_E$, we can define a corresponding Lorentzian geometry by analytically continuing $t_E \to -it$. This has $n$ asymptotically AdS regions, connected by a collapsing wormhole which generalises the Einstein-Rosen bridge in the eternal black hole. Topological censorship implies that associated to each boundary of the geometry is a horizon \cite{Friedman:1993ty,Galloway:1999bp}. The absence of local degrees of freedom implies that the geometry in the exterior regions outside the horizons is exactly the BTZ geometry exterior to a black hole.

We want to understand the structure of the dual CFT state which encodes this geometry, and specifically its entanglement. The holographic description of these geometries was initiated in \cite{Krasnov:2000zq,Krasnov:2003ye,Skenderis:2009ju}. The conformal boundary of this spacetime lies at $t_E \to \pm \infty$, and consists of two copies of the surface $\Sigma$. The CFT path integral over this surface has a rich phase structure \cite{Balasubramanian:2014hda,Maxfield:2016mwh}. In a region of the moduli space, the dominant bulk contribution comes from the multiboundary wormhole \eqref{H3}, where the spatial slices are the Riemann surface $\Sigma$. Thus, in this region of moduli space the $t=0$ bulk geometry $\Sigma$ corresponds to a CFT state on the $n$ boundaries obtained by a path integral on $\Sigma$. 

For the BTZ black hole, the entanglement structure of the state \eqref{TFD} is purely bipartite. In the high temperature limit, this has a particularly simple structure: high temperature is small $\beta$, so the cylinder is short, and if we consider scales larger than the thermal scale $\beta$ on the spatial circle, the path integral simply identifies states on the two boundaries. This local character of the entanglement was verified in \cite{Morrison:2012iz} by considering mutual informations between subregions on the two boundaries.  

\begin{figure}
\centering
\includegraphics[width=0.5\textwidth]{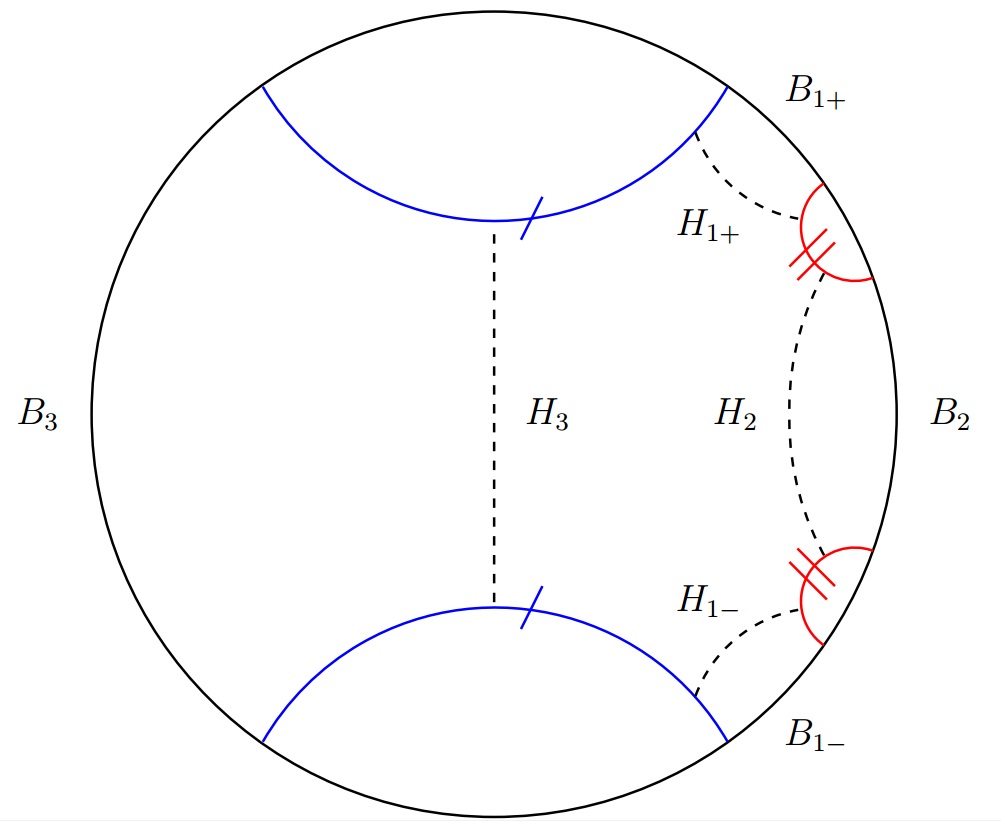}
\caption{ The surface $\Sigma$ as a quotient of the Poincar\'e disc for the pair of pants. The pairs of labeled geodesics (blue and red in colour version) are identified by the action of $\Gamma$. The region of the Poincar\'e disc bounded by these geodesics provides a fundamental domain for the quotient. $B_3$, $B_2$ and $ B_1 = B_{1+}  \cup B_{1-}$ become the desired three circular boundaries. There are corresponding minimal closed geodesics $H_3$, $H_2$ and $H_1 =  H_{1+} \cup  H_{1-}$. The lengths $L_a$ of these geodesics fully characterize the geometry of $\Sigma$. }
\label{pants}
\end{figure}

The next simple example is the three-boundary wormhole or pair of pants, whose Euclidean geometry is obtained by quotienting by a group $\Gamma$ generated by a pair of hyperbolic elements $g_1, g_2$. A fundamental domain of the identification on $H^2$ is the region bounded by a pair of geodesics identified by $g_1$ and a pair of geodesics bounded by $g_2$, as depicted in figure \ref{pants}. This surface has three moduli, corresponding to the lengths of the three minimal closed geodesics shown in the figure. In the Lorentzian spacetime, these geodesics become the bifurcation surfaces of the event horizons in each asymptotic region. The CFT path integral on the pair of pants is hard to do analytically, but it simplifies in limits of the moduli space. In \cite{Balasubramanian:2014hda} the entanglement properties of the dual state were studied in the ``puncture limit'', where the minimal geodesics are short.  

\begin{figure}
\centering
\begin{subfigure}[t]{0.2\textwidth}
\centering
\includegraphics[scale=0.6]{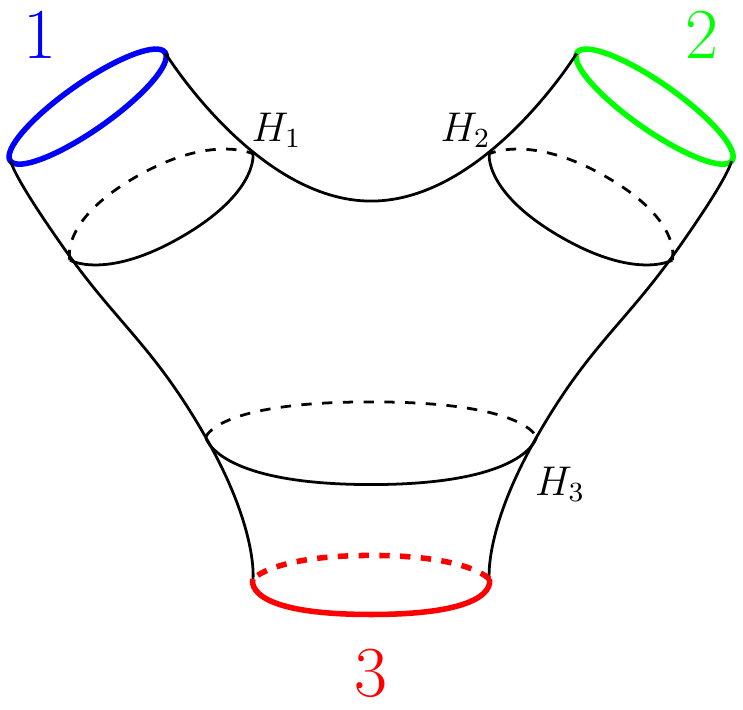}
\caption{}
\end{subfigure}
\hspace{.08\textwidth}
\begin{subfigure}[t]{0.2\textwidth}
\centering
\includegraphics[scale=0.6]{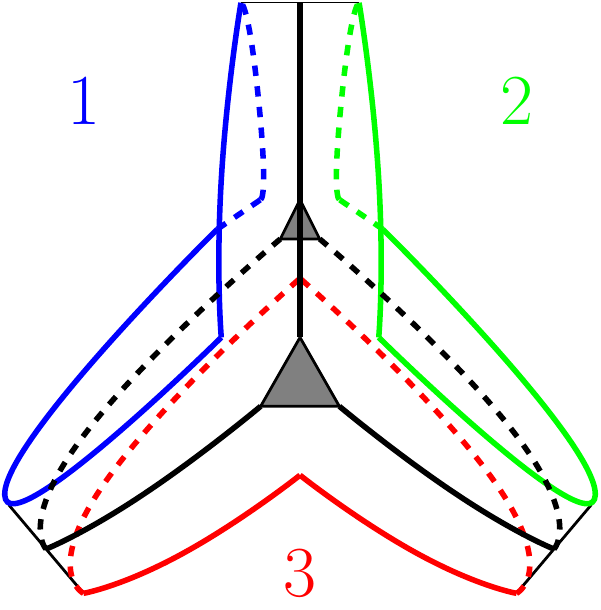}
\caption{}
\end{subfigure}
\begin{subfigure}[t]{0.3\textwidth}
\centering
\includegraphics[scale=0.6]{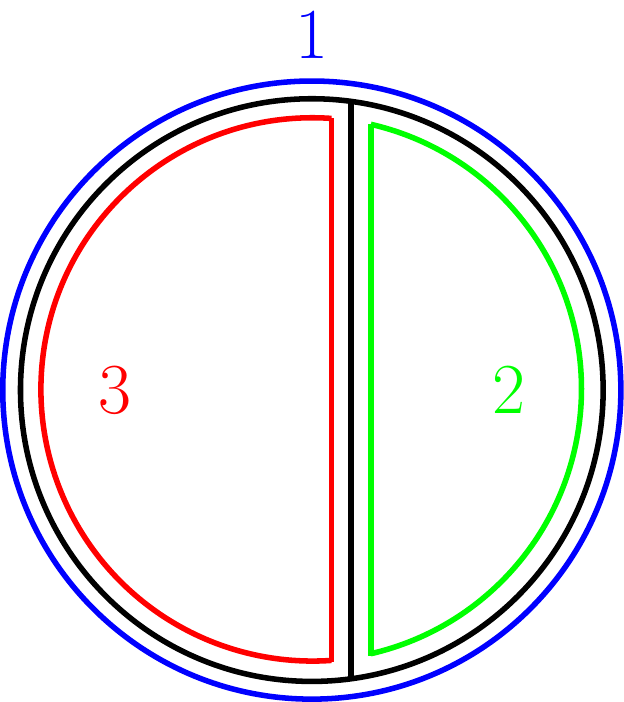}
\caption{}
\end{subfigure}
\caption{The geometry of the pair-of-pants in the high temperature limit and the resulting entanglement structure. a) the pair-of-pants geometry with the three coloured boundaries indicated and labelled 1-3, the black lines depict the horizons pertaining to each exterior region, and are labelled $H_a$, $a=1,2,3$. The interior of the horizons is the causal shadow region. b) Cartoon of a) in the high-temperature limit, with fixed ratios of the moduli. The exterior cylinders shrink (the strips should be thought of as being extremely thin, we've exaggerated them here) and the distance between the horizons across the causal shadow region is small almost everywhere. The black lines represent identifications between horizons, which is true to exponential accuracy away from the junctions.  c) The resulting entanglement structure can be depicted with this ``wheel'' diagram: the path integral locally identifies the states in portions of the three boundaries. States localised in some boundary interval are purified by an interval of the same size on the opposite side of the seam, which may lie on any of the three boundaries, as the ratios of the moduli are varied. The resulting entanglement structure is almost entirely bipartite.
}
\label{3BHT}
\end{figure}

In \cite{Marolf:2015vma}, the structure in the ``high-temperature'' limit, where the geodesics are long, was studied. This leads to particular simplifications. For the three-boundary wormhole, the ``high-temperature'' limit is defined by scaling the sizes of all of the horizons to infinity, whilst fixing their ratios, which then characterise the high temperature geometry. The geometry outside the horizons are high-temperature BTZ solutions, which justifies the name, although the CFT state on the boundaries is not thermal. Since the exterior cylinders are BTZ, they behave in the same way as before: considering scales above the thermal scale, the state on the boundary is identified with the state on the horizon. There is a causal shadow region between the horizons, but its volume is fixed in AdS units by the Gauss-Bonnet theorem, so as the horizons become long, the distance between them shrinks over almost all of the horizon. The causal shadow region is thus effectively a seam which connects the horizons of the exterior regions and whose shape is determined by the ratios of the moduli. Thus, in the high-temperature limit we infer that the path-integral just identifies states across this seam, so that intervals in different boundaries which are opposite each other across the shadow region are maximally entangled, and again the resulting entanglement structure is almost \emph{entirely} bipartite and local. This behaviour is depicted in figure \ref{3BHT}. There could be some residual multipartite component, but this would only involve a subregion of order the thermal scale on each boundary.  

Note the figure depicts the regime where the horizons are all roughly of the same length. If we take $L_1 \geq L_2 \geq L_3$, this is the regime $L_1 < L_2 + L_3$, referred to in \cite{Marolf:2015vma} as the ``wheel" regime, after the figure on the right side of figure \ref{3BHT}. The alternative regime  $L_1 \geq L_2 + L_3$ is referred to as the ``eyeglass" regime. The entanglement remains primarily bipartite in this regime, but there are regions of boundary 1 which are entangled with other regions of boundary 1, rather than with one of the other boundaries. 

The result generalises easily to wormholes with more boundaries and topology behind the horizon. Any Riemann surface can be decomposed into pairs of pants, sewn together across minimal geodesics. There is a region of the moduli space where all the minimal geodesics involved in the sewing are long, and the individual pairs of pants are in the high-temperature configuration described above. The path-integral then identifies states across the regions between minimal geodesics, again generating a local bipartite entanglement structure which can be characterised by appropriate compositions of the diagrams of which figure \ref{3BHT} $c)$ is an example.

\section{Hyperbolic Tilings \& Quotients}
\label{tiling}

The tensor network models of \cite{Pastawski:2015qua,Hayden:2016cfa} are based on tiling the hyperbolic plane with perfect or random tensors. We want to take a quotient of these networks by a discrete isometry of the network to obtain a model of the multiboundary wormholes. We can usefully seperate the geometrical aspects of choosing a tiling of the hyperbolic plane and its quotients by discrete isometries from the choice of tensors, so we will first discuss the geometric aspects in this section. 

Introducing a regular tiling of the hyperbolic plane provides us with a natural discretization of $H^2$. A particular choice of tiling will preserve some discrete subgroup of the $SL(2,\mathbb{R})$ isometries of $H^2$, and we can quotient by some of these isometries of the tiling to obtain discretizations of the Riemann surface $\Sigma$ for some discrete values of the moduli. In this section, we describe the tilings and quotients, and define analogues of the horizons in the tiling. 

\begin{figure}
\centering
\includegraphics[width=0.7\textwidth]{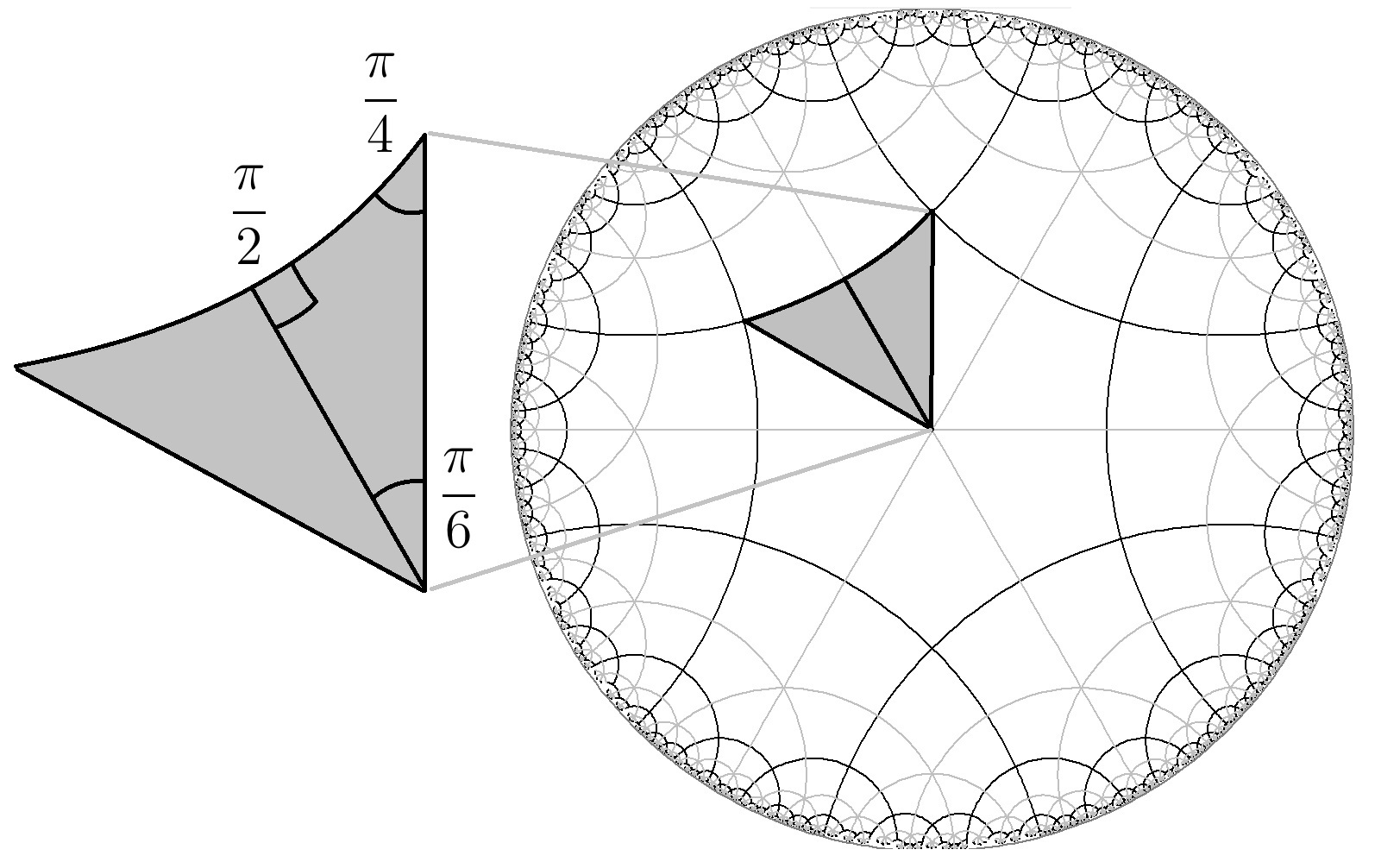}
\caption{The [2,4,6] tiling of the hyperbolic plane $H^2$. One of the seed triangles is indicated, the whole tiling is covered by the action of reflecting the seed along its edges. The black, primary tiling lines divide the network into a regular array of hexagons.}
\label{246}
\end{figure}

As in \cite{Bhattacharyya:2016hbx}, we describe the tilings in terms of Coxeter groups \cite{Coxeter:na, Murray:na, Wiki:na}. A constant curvature connected Riemann surface can be tiled by repeated reflections of a seed triangle about its edges. If the interior angles of the seed triangle are given by $\frac{\pi}{r}$, $\frac{\pi}{p}$ and $\frac{\pi}{q}$ for $p,q,r \in \mathbb{Z}^+$ then the set of all reflections in its edges form a Coxeter group, denoted $[r,p,q]$. The triangulation of a space obtained by repeated reflection of a seed triangle in its edges is referred to as a Coxeter tiling, and the Coxeter group forms the discrete isometries of the tiling. By the Gauss-Bonnet theorem, for hyperbolic spaces, the required seed triangle must have
\begin{equation}
\label{gaussbonnet}
\frac{1}{p}+\frac{1}{q}+\frac{1}{r} < 1.
\end{equation}
The $[r,p,q]$ tilings satisfying \eqref{gaussbonnet} tile $H^2$ with a regular array of $q$-gons. These Coxeter tilings underly the tensor networks considered in \cite{Pastawski:2015qua,Hayden:2016cfa,Bhattacharyya:2016hbx}. An example hyperbolic tiling is illustrated in figure \ref{246}. The edges of the $q$-gons are geodesics in $H^2$, so the volume of each $q$-gon is of order the AdS scale. 

The tensor networks considered in \cite{Pastawski:2015qua,Yang:2015uoa,Hayden:2016cfa} are constructed by thinking of the tiling as a graph, and taking the tensor network to be the dual graph. That is, each $q$-gon face in the tiling is replaced by a vertex in the tensor network, which has $q$ legs,  connecting it to the tensors in the adjacent faces, across the edges of the tiling. There are uncontracted legs at the boundary of the hyperbolic plane.

The tiling is invariant under reflections in any of these edges, forming the Coxeter group associated with the tiling. We can quotient by any subgroup of this group. In the construction of the Riemann surface $\Sigma$ in the previous section, we considered quotients by hyperbolic elements, which identified pairs of geodesics in the hyperbolic plane. We can obtain such hyperbolic elements by combining a pair of reflections in distinct geodesics \cite{Bhattacharyya:2016hbx}. We can see that this follows by considering $AdS_3$ as the hyperboloid embedded in $\mathbb{R}_4$,
\begin{equation}
\label{r4}
X_0^2 + X_3^3 - X_1 ^2 - X_2^2 = l^2
\end{equation}
the $t=0$ slice corresponds to $X_3=0$. Now consider the two hyperplanes $P_1$ and $P_2$ with corresponding normals $n_1 = (0,0,1,0)$, $n_2 = (\cosh(\frac{\eta}{2}), \sinh(\frac{\eta}{2}), 0, 0 )$ respectively. Under a reflection in the hyperplane with normal $n$, a point $X$ transforms as
\begin{equation}
\label{reflection}
\tilde{X} = X - 2 (n \cdot X) n
\end{equation} 
So that under the pair of subsequent reflections, first in $P_1$ and then in $P_2$, the $t=0$ slice is preserved whilst the $X_0$ and $X_2$ are Lorentz boosted, by 
\begin{equation}
\label{lt}
\left(
\begin{array}{c}
\tilde{X_0}\\
\tilde{X_2}\\ 
\end{array}
\right)
=
\left(
\begin{matrix}
    -\cosh(\eta) & \sinh(\eta) \\
    \cosh(\eta) & -\sinh(\eta) \\
\end{matrix}
\right)
\left(
\begin{array}{c}
X_0\\
X_2\\ 
\end{array}
\right).
\end{equation}

\begin{figure}
\centering
\includegraphics[width=0.7\textwidth]{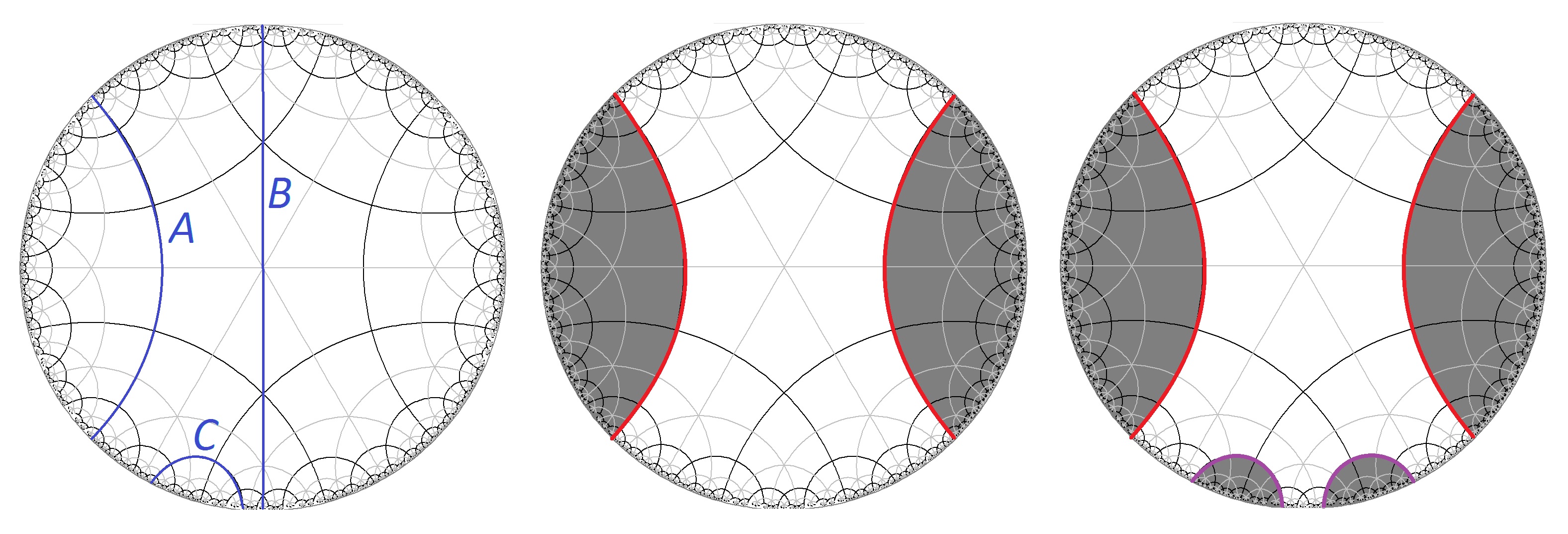}
\caption{An illustration of the quotient operation in the [2,4,6] tiling. We take $r_A$, $r_B$, $r_C$ to be the reflections in the geodesics labelled in the left-hand diagram. Quotienting by $\Gamma$ generated by $g_1 = r_A r_B$ gives the tiling of BTZ shown in the middle diagram. Quotienting by $\Gamma$ generated by $g_1 = r_A r_B$ and $g_2 = r_B r_C$ gives the tiling of the pair of pants shown in the right-hand diagram. The unshaded region is a fundamental region for the identification in both cases.}
\label{quotients}
\end{figure}

Thus, the hyperbolic element identifying any pair of geodesics which form edges of the tiles will be an isometry of the tiling. We can thus quotient by discrete groups $\Gamma$ composed of such hyperbolic elements to obtain a tiling of a Riemann surface $\Sigma = H^2/\Gamma$, and hence (considering the dual graph) a tensor network with the topological structure of $\Sigma$.  We can construct this tiling only for some discrete choices of the moduli of the Riemann surface, as given a tiling, the area of the tiles, and hence the distance between the geodesics to be identified, is fixed by the Gauss-Bonnet theorem. In \cite{Bhattacharyya:2016hbx}, this approach was used to obtain tilings and hence tensor networks corresponding to the BTZ geometry. We now want to generalize this to multiboundary wormholes. Our detailed analysis will focus mainly on the pair of pants. An illustrative example of the construction is given in figure \ref{quotients}.

\begin{figure}
\centering
\includegraphics[width=1\textwidth]{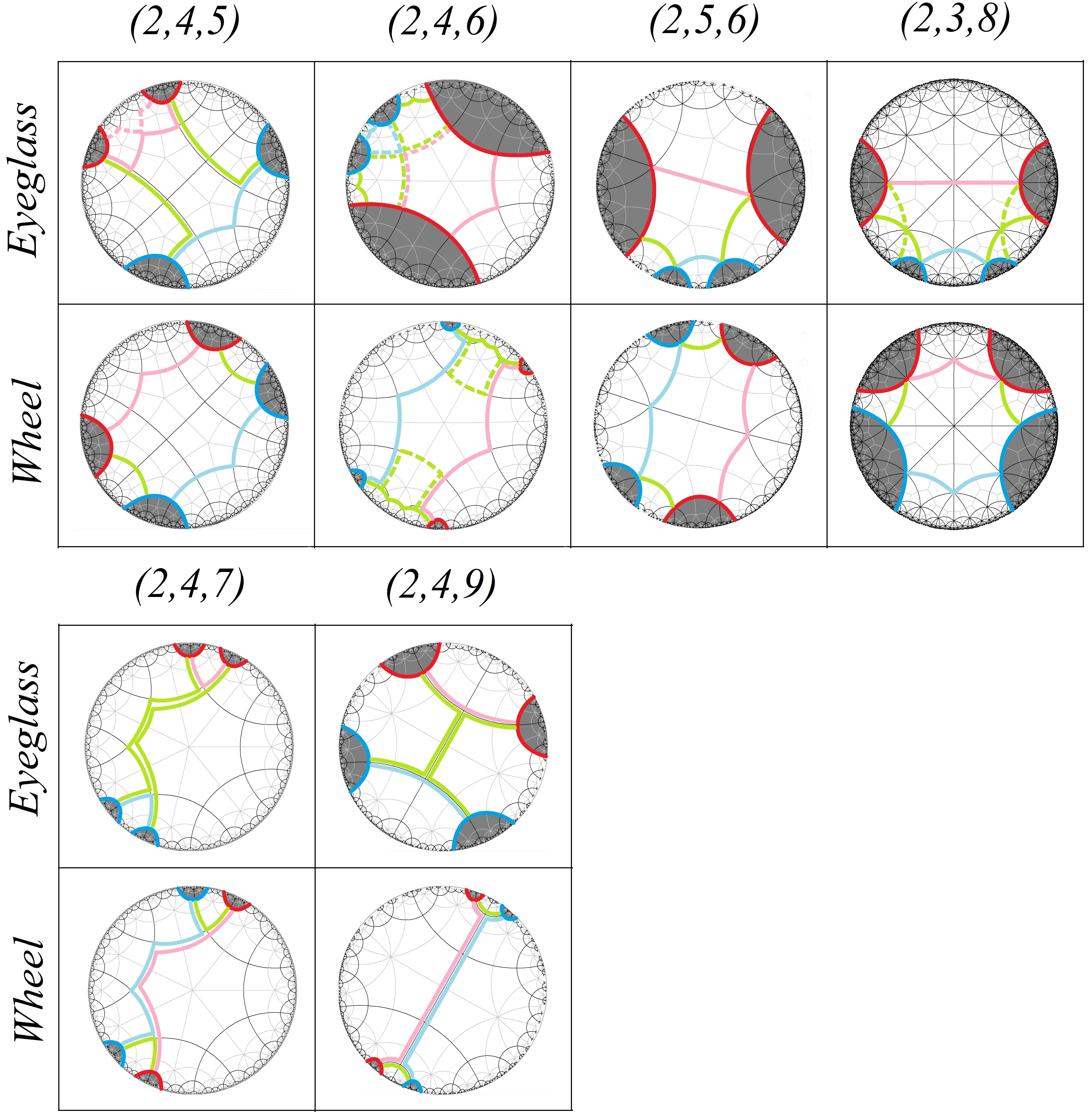}
\caption{Illustrations of three-boundary tilings obtained by quotienting the tilings of the hyperbolic plane indicated by each column, and with discrete moduli in the regimes indicated by each row. The unshaded region is a fundamental region for the identification. The minimal closed paths along tile boundaries homologous to each conformal boundary are indicated. The cases with two paths of the same colour, where one is dashed, represent degeneracies in the choice of closed path. The second group of examples show cases where there is no causal shadow; the minimal closed paths coincide, so the entanglement between different regions is entirely bipartite.}
\label{3s}
\end{figure}

Despite the discretization of the moduli of $\Sigma$, the minimal geodesics will not generally lie along edges of the tiling. Thus, it is important to identify the analogues in the tiling of these minimal geodesics. We will take this to be the minimal closed path along the edges of the tiling homologous to each boundary. This is natural because once we introduce the tensors in each tiling, this path will cut across the links between tensors.\footnote{Note that this is slightly different from the prescription in \cite{Bhattacharyya:2016hbx}, where the BTZ horizon in the example in figure \ref{quotients} was identified with a tensor in the network, as the actual minimal geodesic runs along the middle of the tile. For the more general case we consider, it is more natural to take the definition above, even though this leads to an artificial degeneracy in the BTZ case.} This will lead to degeneracy in some cases, where there can be multiple paths of the same length along the edges. Some examples are illustrated in figure \ref{3s}.

An interesting feature is that in some cases there is then {\it no} causal shadow region in the tiling; the minimal length paths in the network can coincide. We will see below that this leads to tensor networks where the entanglement is entirely bipartite. In the continuum, there is of course still a causal shadow for these choices of $\Sigma$, which partially covers some tiles, so this could be viewed as just a discretization error, but we argue that it is actually an interesting feature. It implies that in the context of the tensor network models, it is possible to have a network on the pair of pants that gives rise to a state with only bipartite entanglement, providing further evidence that multipartite entanglement is not an essential component in obtaining multiply connected geometries. These explicit examples help us to understand how the geometry arises from purely bipartite entanglement.  

\section{Tensor Networks \& Holography}
\label{networks}

 We now turn to the specific tensor network models we use, following \cite{Pastawski:2015qua,Hayden:2016cfa}. We obtain a network from the tiling by considering the dual graph, with a network vertex in each tile and legs connecting the vertices in adjacent tiles. In general, a network is used to define a quantum state by first associating a tensor $T_{i_1 ... i_n}$ to each vertex in the network, with the rank of the tensor equal to the number of legs at the vertex. We then associate a state $\ket{T}$ in a tensor product Hilbert space $\mathcal{H} =  \otimes_n \mathcal{H}_n$ with the tensor,
\begin{equation}
\label{tndefinition}
\ket{T} =  \sum_{{i_k}} T_{i_1 ... i_n} \ket{i_1}_1 \otimes ... \otimes \ket{i_n}_n,
\end{equation}
where $ \ket{i_k}_k$ is a basis for the $k$th factor with $i_k = 1, \ldots, D_k$ where $D_k$ is the dimension of $\mathcal{H}_k$, which we will take to be some constant $\chi$ for all legs, called the bond dimension. 
Taking the product over all the vertices defines a product state $\ket{ \{ T \}} = \otimes_V \ket{T^V}$. For each leg joining two vertices, we make a projection onto the maximally entangled state
\begin{equation}
\label{projector}
\ket{i j } = \frac{1}{\sqrt{\chi}}\sum_{a = 1}^\chi \ket{a}_{i} \otimes  \ket{a}_{j}
\end{equation}
in the associated Hilbert spaces. This defines a state in the Hilbert space associated to the uncontracted legs,  
\begin{equation}
\label{tncomplete}
\ket{\tilde{T}} = \otimes_{ \{ i j \} } \braket{i j }{ \left\{ T \right\} },
\end{equation}
where the product runs over all the contracted legs, which could include self-contractions in general.

If we take the network constructed from the dual graph of a Coxeter tiling of $H^2$,  the remaining uncontracted legs are located at the boundary of the hyperbolic plane. Given a choice of tensor at each vertex, this defines a state on the boundary legs. This is referred to as a \textbf{holographic state}. In an alternative construction, a $q+1$ legged tensor is associated to each vertex, leaving one uncontracted leg at each vertex in the network, which are referred to as bulk legs, in addition to the uncontracted boundary legs.     This network can be viewed as a map from the Hilbert space $\mathcal{H}_{\text{Bulk}}$ of the bulk legs to the Hilbert space $\mathcal{H}_{\text{Boundary}}$ of the boundary legs. This model can be used to study the encoding of local bulk operators in the boundary Hilbert space, so it is referred to as a \textbf{holographic code}. 

Holographic states realise a discrete version of the Ryu-Takayanagi formula \cite{Ryu:2006bv}, relating the entanglement entropy of some subset of the boundary legs to the length of a cut in the bulk. Suppose we have two boundary regions $A$ and $A^C$, and a cut $\gamma_A$ in the bulk, which is a path in the bulk along edges of the tiling, which cuts through tensor legs, separating the network into two components, such that one component has boundary $\gamma_A \cup A$ and the other has boundary $\gamma_A \cup A^C$. Then the number of legs $|\gamma_A|$ along the cut provides an upper bound for the entanglement entropy of the reduced density matrix on $A$ in the holographic state given by the network \cite{Pastawski:2015qua}:
\begin{equation} \label{nrt}
S_A \leq |\gamma_A| \ln \chi, 
\end{equation}
where $\chi$ is the bond dimension. We obtain the tightest bound by considering the minimal cut, which we can think of as the network analogue of a minimal surface. This bound is saturated if the two components of the network are isometries from $\gamma_A$ to $A$ and from $\gamma_A$ to $A^C$. This then realises a lattice version of Ryu-Takayanagi, relating the entropy to the length of the minimum cut. There can be degenerate minimal cuts in the networks, though this does not alter the bound \ref{nrt} as in such cases, the minimal cuts are of equal length.

\begin{figure}
\centering
\includegraphics[width=0.9\textwidth]{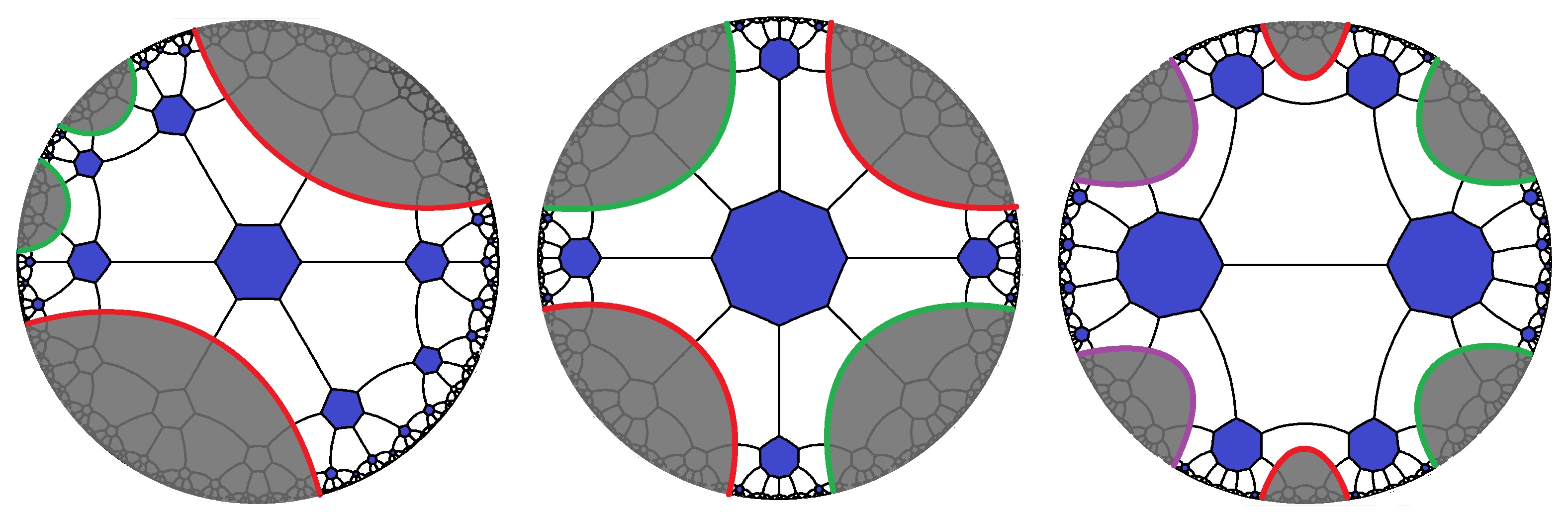}
\caption{Some examples of topologically interesting holographic states constructed from quotients of various coxeter tilings. The fundamental domain between colour-coded pairs of identified cuts is the unshaded region in each case.  These are holographic states representing the three-boundary wormhole (left), a torus wormhole (centre) and a four-boundary wormhole (right).}
\label{zoo}
\end{figure}

Applying the same prescription to the quotient tilings, we can build tensor networks on a Riemann surface $\Sigma$, giving states and codes for multiboundary geometries. A selection of examples are illustrated in figure \ref{zoo}.

\subsection{Perfect Tensors}

\begin{figure}
\centering
\includegraphics[width=0.5\textwidth]{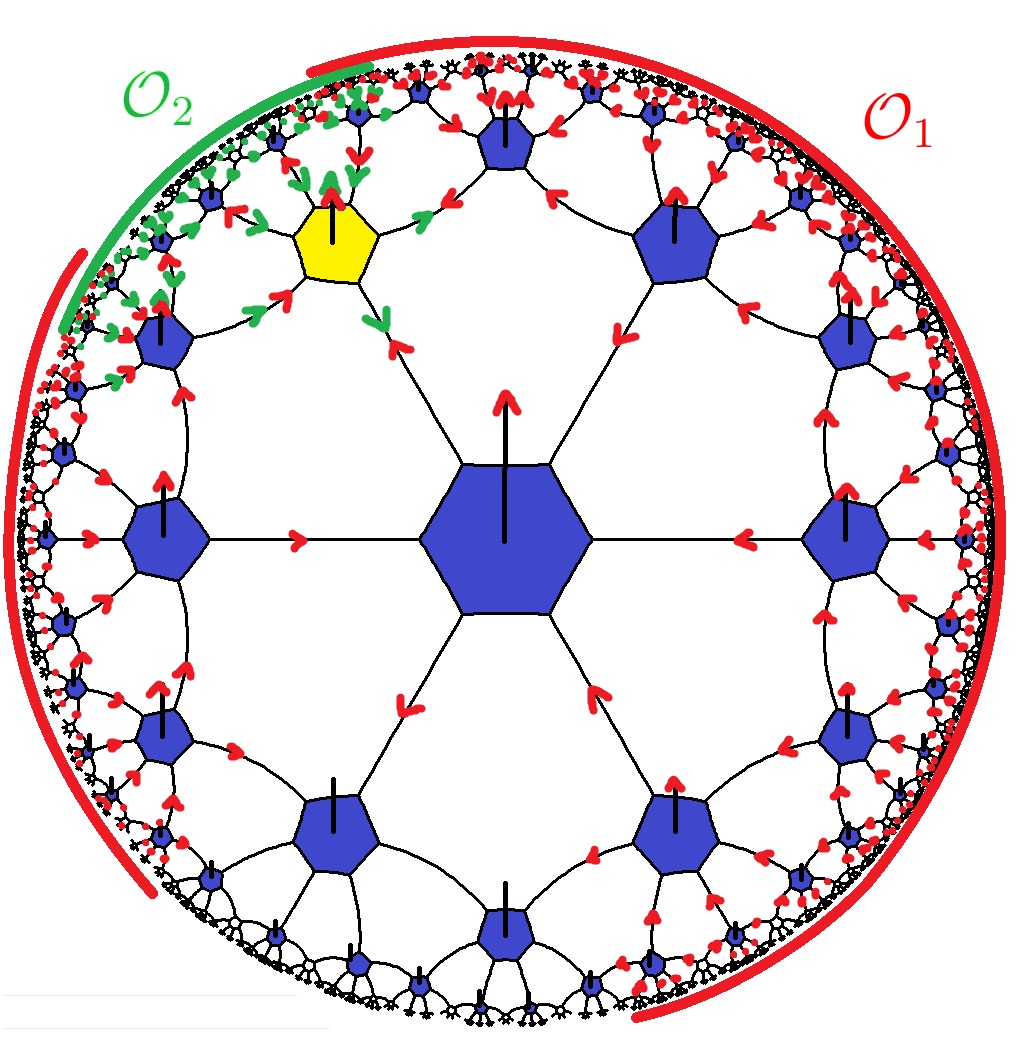}
\caption{Bulk reconstruction for the holographic heptagon code. Due to the fact that each tensor is perfect, the local bulk operator $\mathcal{O}$ acting on the bulk leg of the yellow-highlighted tensor can be pushed to a non-local boundary operator in multiple ways. Two examples are indicated, whereby the operator is pushed, on the one-hand, to the operator $\mathcal{O}_2$ acting on the connected interval shown in green, and on the other hand to the operator $\mathcal{O}_1$ acting on the disconnected interval shown in red. At each vertex, the tensor is proportional to an isometry from the outgoing legs to the ingoing legs, so we can push the operator back from the outgoing legs to the ingoing legs.}
\label{qecads}
\end{figure}

In \cite{Pastawski:2015qua}, the tensors at each vertex were taken to be perfect tensors. A perfect tensor is a $2n$ index tensor $T$ such that for any division of its indices into a set $A$ and its complement $A^C$ such that $Dim(\mathcal{H}_{A}) < Dim(\mathcal{H}_{A^C})$, $T$ is proportional to an isometry from $\mathcal{H}_{A}$ to $\mathcal{H}_{A^C}$. That is, the map from $\mathcal{H}_{A}$ to $\mathcal{H}_{A^C}$ preserves the inner product up to an overall factor. If we denote the indices in $A$ by a collective index $a$, and the indices in $A^C$ by $b'$, the condition is 
\begin{equation}
\sum_{b'} T^\dagger_{a b'}T_{b' c} = C \delta_{ac} 
\end{equation}
for some constant $C$. The isometry property implies that we can convert an operator acting on $\mathcal{H}_{A}$ into an operator acting on $\mathcal{H}_{A^C}$; given an operator $\mathcal{O}$ acting in $\mathcal{H}_A$, we define 
\begin{equation}
\tilde{\mathcal{O}} = \frac{1}{C} T \mathcal{O} T^\dag
\end{equation}
acting in $\mathcal{H}_{A^C}$, so that $T \mathcal O = \tilde{\mathcal O} T$. In a holographic code, this enables us to rewrite an operator acting on a bulk leg as an operator acting on some subspace of the boundary Hilbert space, by using the perfect tensors in the network to push the operator outwards, as illustrated in figure \ref{qecads}. This provides a tensor network realisation of bulk reconstruction. Since we can use the perfect tensor property to map the bulk leg to different subsets of the boundary legs, it can be mapped to operators acting on different subspaces of the boundary Hilbert space, realising the ideas of \cite{Almheiri:2014lwa}. 

In \cite{Pastawski:2015qua}, a greedy algorithm was introduced to identify the portion of the bulk that can be reconstructed from a given region $A$ of the boundary, not necessarily connected. This proceeds by taking some initial region that can be reconstructed from the boundary region (consisting of tensors in the asymptotic region) and iteratively adding to this region a tensor with more than half of its legs connected to tensors already in the region, until there are no more such tensors. The boundary of this region is a cut of the network referred to as the greedy geodesic $\gamma_A$. In the case of the holographic state, the collection of tensors $\mathcal{G}_A$ lying between $A$ and $\gamma_A$ define an isometry from $\gamma_A$ to $A$. For holographic codes, we have an isometry from $\gamma_A$ and the bulk Hilbert space in $\mathcal{G}_A$ to $A$. This gives a tensor network realisation of the idea of the bulk wedge associated with a given boundary region. Since $\mathcal{G}_A$ defines an isometry, we can view moving from the boundary region $A$ to $\gamma_A$ as a process of distillation, extracting the degrees of freedom in $A$ which are entangled with $A^C$. 

If we divide the boundary into several different regions, there will be a greedy geodesic associated to each of them, and the union of the different wedges $\mathcal{G}_A$ may not cover the whole network. The remaining portion was called in \cite{Pastawski:2015qua} the residual multipartite region, and can be thought of as encoding the entanglement between the different boundary regions. We expect the causal shadow region in our quotient networks to play a similar role. 

It was also shown in \cite{Pastawski:2015qua} that for holographic states, for any connected region $A$ on the boundary of a simply-connected perfect tensor network of non-positive curvature, the lower bound in \eqref{nrt} is saturated, so that the lattice Ryu-Takayanagi formula holds. 

\subsection{Random Tensors}

In \cite{Hayden:2016cfa}, a different approach was taken based on selecting the individual tensors in the network independently at random from a suitable distribution. This corresponds to taking the state at the individual vertices \eqref{tndefinition} to be a Haar random state. 

In the limit of large bond dimension $\chi$, the calculation  of the second Renyi entropy averaged over the randomness was mapped to a partition function of an Ising spin system. This was used to show that these random tensor networks also satisfy a lattice Ryu-Takayanagi formula. 

Random tensors are not perfect tensors, but it was shown that in the limit of large bond dimension, they are approximately perfect tensors. This is essentially because the Page theorem \cite{Page:1993df} says that a random tensor is \emph{approximately} an isometry from any subset of \emph{less} than half of its indices to the remaining set. The maximum entanglement entropy of the reduced density matrix on an $n$-dimensional share of a state in an $(m+n)$-dimensional Hilbert space is
\begin{equation}
\label{page}
S_{\text{max}} = Log(m) - \frac{m}{2n} + \text{...}
\end{equation}
where the ``...'' terms refer to terms subleading in $\frac{m}{n}$. For a Haar random state of dimension $n = \chi N$, defined by an $N$-legged tensor $T$ with bond dimension $\chi$, \eqref{page} implies that the reduced density matrix on any subset of $M < \frac{N}{2}$ legs is approximately maximally mixed and hence the map from the $M$ legs to the remaining $(N-M)$ legs is an approximate isometry. This approximate isometry is sufficient for results similar to the perfect tensors to apply; we can map local bulk operators in a holographic code to operators acting on subspaces of the boundary, and given a boundary region there is a corresponding bulk region which is reconstructable. 

\section{Multiboundary Networks}
\label{multi}

\subsection{BTZ}

We now apply these tensor network constructions to the tilings obtained for multiboundary geometries in section \ref{tiling}. If we consider first the BTZ black hole, there is a single identification. Asymptotically far from the black hole, the network will look like the network for $H^2$. The region outside a large black hole was already analysed qualitatively in \cite{Pastawski:2015qua}. For holographic states, the network lying between the minimal closed path and the boundary will be an (approximate) isometry, so we can think of the network legs lying across the minimal closed path as representing the distilled entanglement between the two asymptotic regions. This minimal closed path provides a minimal cut in the network, where we take the region $A$ to be the whole of one of the boundaries. In this case the entanglement is entirely bipartite, with each leg across the cut corresponding to a pair of Hilbert spaces in the maximally entangled state \eqref{projector} - the analogue of a Bell pair for systems of dimension $\chi$. 

\begin{figure}
\centering
\includegraphics[width=\textwidth]{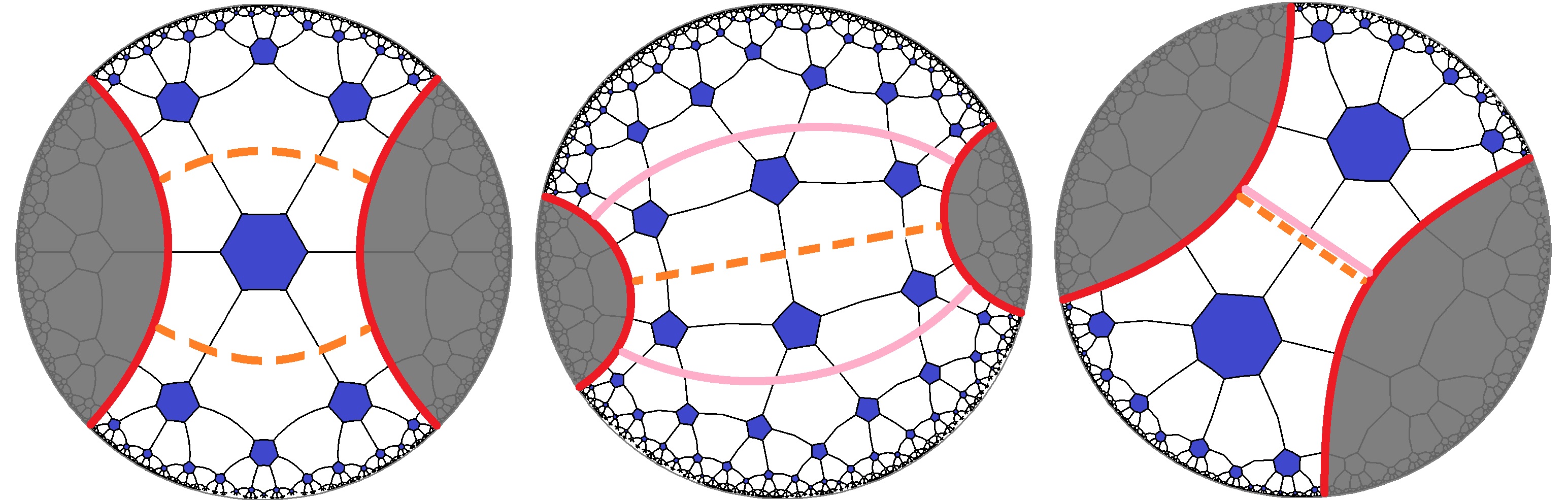}
\caption{BTZ networks exhibiting tiling artifacts. The network shown on the left has two equal-length minimal cuts, being \emph{either} of the orange, dashed cuts. The continuum horizon is not a mirror of the tiling underlying this network. The network in the centre has a unique minimal cut, but has a bipartite residual region consisting of the 6 tensors lying inbetween the pink greedy geodesic (being a pair of disconnected cuts). Due to the identification, there are too few legs crossing the greedy geodesic to be able to push it to the minimal cut. The network on the right exhibits the behaviour we expect; there is a unique minimal cut that is reached by the greedy geodesic so that we can distill all of the entanglement between the two boundaries to Bell pairs crossing the minimal cut.}
\label{dbtz}
\end{figure}

Our choice of a minimal closed path as the analogue of the horizon introduces some minor differences from the analysis of \cite{Bhattacharyya:2016hbx}. As previously noted, some choices of BTZ tiling lead to degenerate minimal closed paths, as in the left example in figure \ref{dbtz}. This is just a failure of the discretization; the actual minimal geodesic does not lie along the tile boundaries, so there are degenerate approximations to it. 

The greedy algorithm can also fail to reach all the way to the minimal closed path, so the greedy geodesic associated to one boundary may be different from the minimal closed path, leading to the appearance of a non-trivial bipartite residual region, as depicted in the central example in figure \ref{dbtz}. The failure of the greedy geodesic to reach the minimal closed path arises because of closed loops in the network, where the number of legs pointing ``out" towards the boundary is smaller than the number of other legs. Since we haven't yet reached the minimal closed path, the number of legs pointing ``out" must overall be less than the number of legs pointing ``in", but the legs around the identification can lead to situations where the exterior legs do not contain enough information to reconstruct the tensors in the shadow region.   

\begin{figure}
\centering
\includegraphics[width=\textwidth]{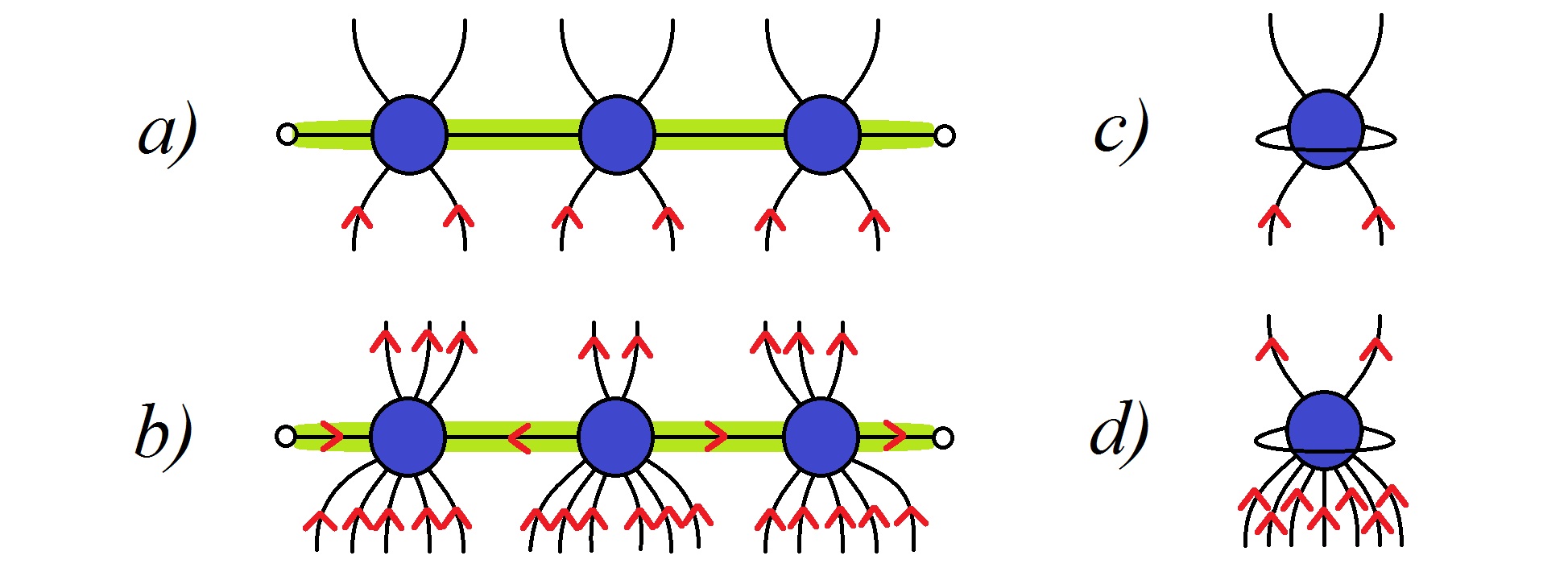}
\caption{Reconstruction with loops and self-contractions within tensor networks. for a) there is not enough information on the lower legs for the greedy algorithm to push across the loop (green), in either the random or perfect tensor case. b) There is enough information on the lower legs of the central tensor for the greedy algorithm to push onto the loop using the arrow assignment depicted. c) The self-contracted tensor is not an isometry from the upper to the lower legs. d) there are sufficiently many legs below for the self-contracted tensor to be an approximate isometry from the upper to the lower legs in the large $\chi$ limit.}
\label{loop}
\end{figure}

Self-contractions of tensors within the network are an interesting special case which are a new feature of the quotient networks. For the perfect tensors, the self-contracted tensor is no longer an isometry from a subset of the remaining legs to the other legs. For sufficiently generic perfect tensors, it is however still an approximate isometry from less than half the remaining legs to the other subset of remaining legs. This feature can inhibit bulk reconstruction by preventing greedy geodesics from crossing loops in order to reach a minimal cut, as depicted in figure \ref{loop}. 

For closed loops, each tensor remains a perfect tensor, but if there are the same number of legs below and above a loop, there will not be enough information in the legs below the loop to be able to push onto the loop itself and subsequently be able to reconstruct the outgoing legs, as depicted in figure \ref{loop} a). If there are more legs below, we can push onto the loop and push past it, as in figure \ref{loop} b). 

\subsection{Multiboundary wormholes}

Considering a more general Riemann surface $\Sigma$, the portion of the network between a given boundary and the minimal cut homologous to this boundary is the same as for BTZ; thus, this defines an isometry from the minimal cut to the boundary, and we can distill the state on the boundary to a state on the legs crossing the cut, which describes the entanglement with the other regions. Hence, the causal shadow region lying between these minimal cuts encodes the entanglement between the different boundaries. For the three boundary case, this causal shadow region will encode the residual tripartite entanglement between the three boundaries. 

\begin{figure}
\centering
\includegraphics[width=0.6\textwidth]{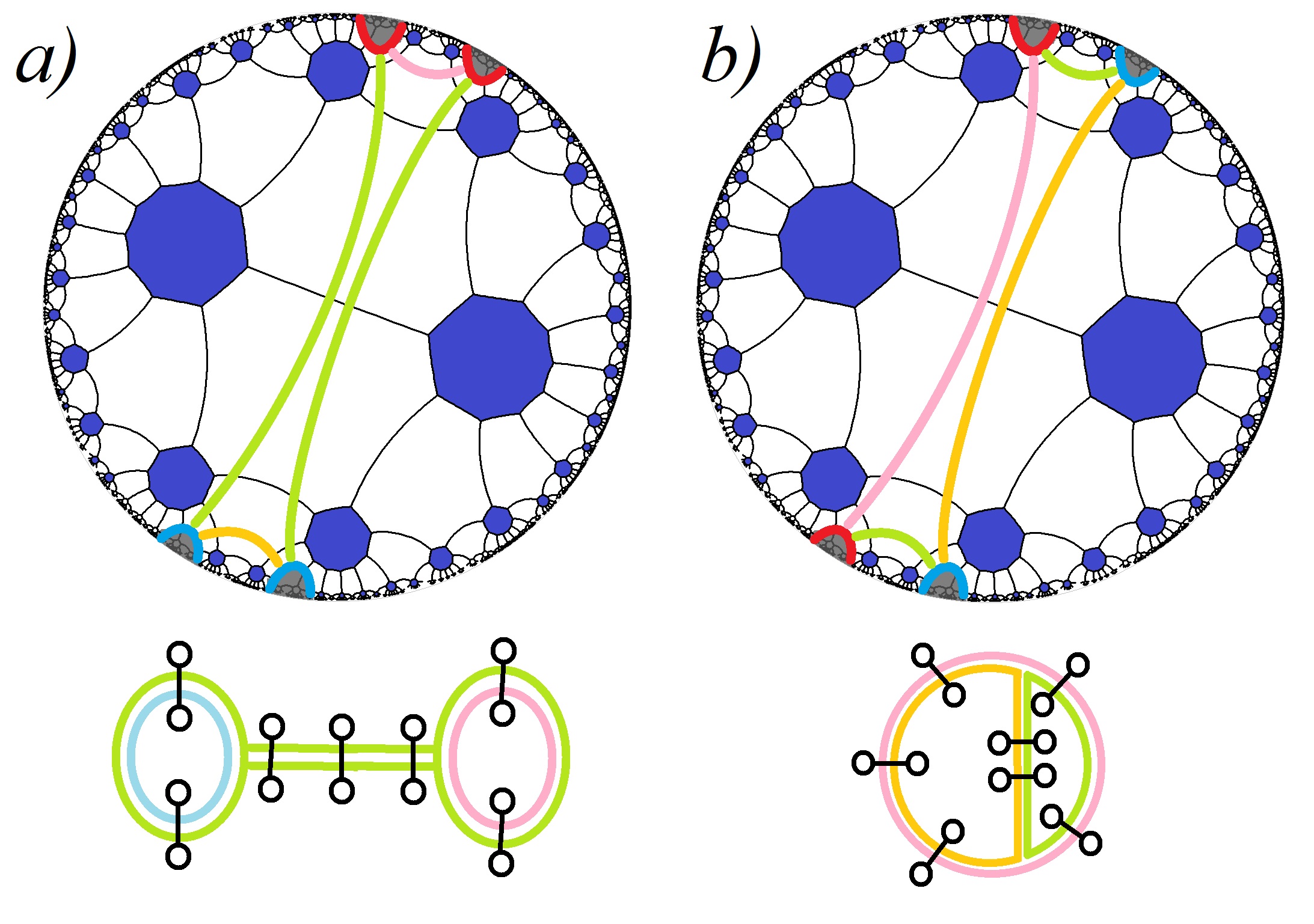}
\caption{Purely bipartite tensor network states for low-T three-boundary wormholes. The fundamental domain is the unshaded region, and the minimal cuts homologous to each region are colour-coded. In each case, there is no multiparty residual region, and states across the minimal cuts are identified. The set of Bell pairs corresponding to states identified across each cut are depicted below each network; coloured lines guide the eye to recognise across which minimal cuts states are identified. These reproduce, schematically, the known structure of the high-T entanglement.}
\label{biponly}
\end{figure}

As noted in section \ref{tiling}, we can have examples where there is no causal shadow region. In this case, we can now see that the entanglement in the holographic state is entirely bipartite, encoded in the maximally entangled states \eqref{projector} on each leg crossing the cuts. This is similar to the entanglement structure seen in the high temperature limit of the CFT path integral, but it is surprising that we can find cases where the entanglement is {\it entirely} bipartite, and that this can occur even for generic moduli. Examples of this behaviour for the wheel and eyeglass regimes are illustrated in figure \ref{biponly}.

\begin{figure}
\centering
\includegraphics[width=0.8\textwidth]{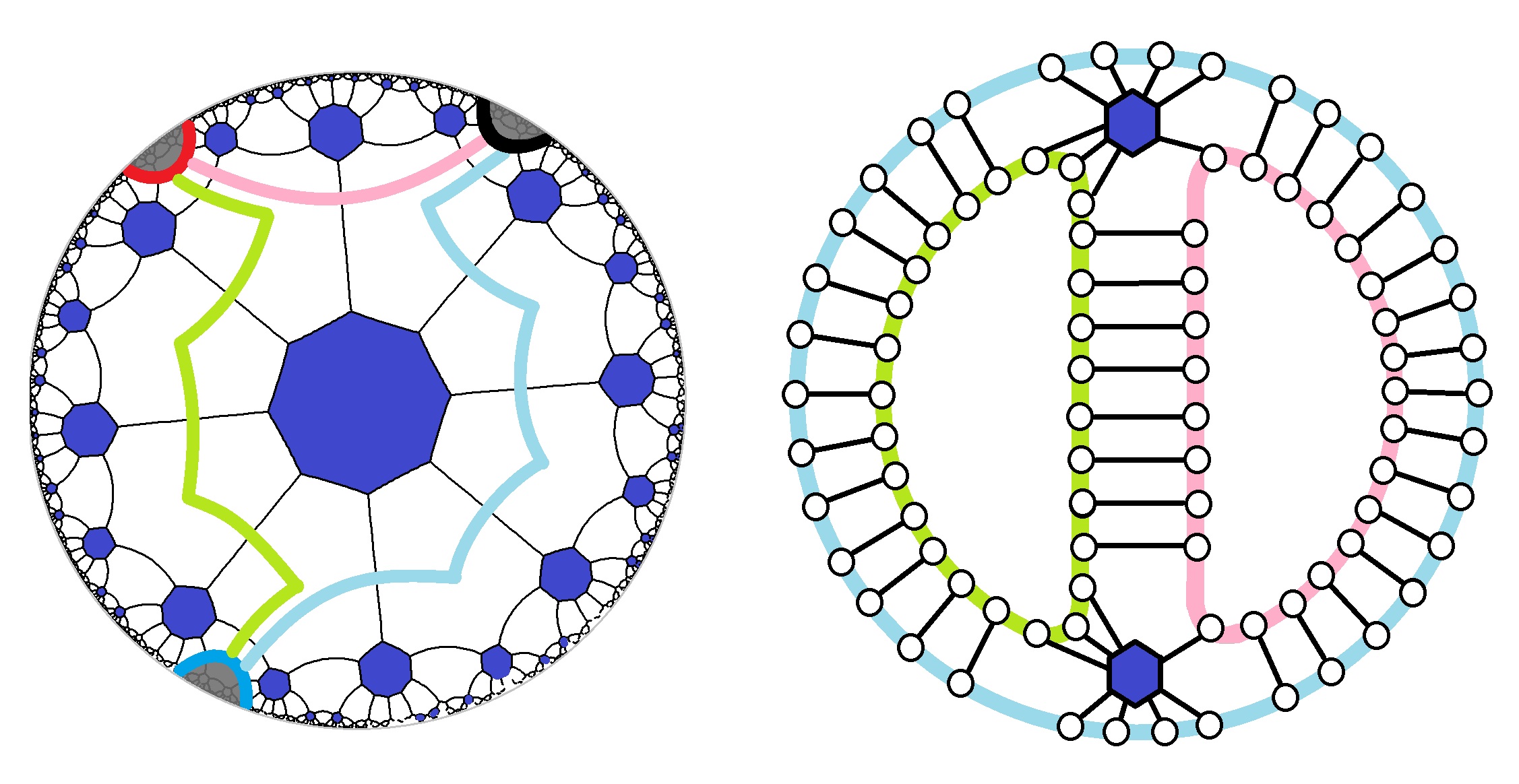}
\caption{Taking the high-T limit of a low-T three-boundary network. For convenience, only half of the tiling (left) is displayed, the non-visible half (the mirror image of what is depicted) is located on the opposite side of the black mirror. We can take a high-T limit by letting the red, blue and black mirrors retreat, increasing the length of each horizon by the same amount. This produces a wheel network in which the horizons have approximately equal lengths. Notably, the tripartite residual region representing the causal shadow, being the central tensor (along with it's unseen reflection) between the coloured cuts, is invariant in the limit. This accords with our intuition that the size of the causal shadow is fixed due to the Gauss-Bonnet theorem. As the minimal cuts become larger, they become identified for most of their length, giving rise to the entanglement structure depicted in the cartoon (right). Bell pairs are identified across the minimal cuts, up to the pair tensors depicted, inhabiting the tripartite residual region.}
\label{htscaling}
\end{figure}

In other cases, there will be a tripartite residual region, and we can ask about the importance of this region and the nature of its entanglement. It is interesting to first make contact with our previous work in the high-temperature limit. We can do so by carefully choosing a low-T network and then letting the quotient mirrors retreat to produce a high-T network with large horizons. What we expect to find in this limit is that the minimal cuts associated to each of the three boundaries become identified for most of their length up to an $AdS$ scale tripartite residual region consisting of only a small number of tensors, whose size remains fixed in this limit. One example of this behaviour is illustrated in figure \ref{htscaling}. Note however that it requires a choice of network to realise these expected features; in other cases the causal shadow grows (or disappears altogether) due to the tiling artefacts illustrated in figure \ref{dbtz}.  

\begin{figure}
\centering
\includegraphics[width=0.7\textwidth]{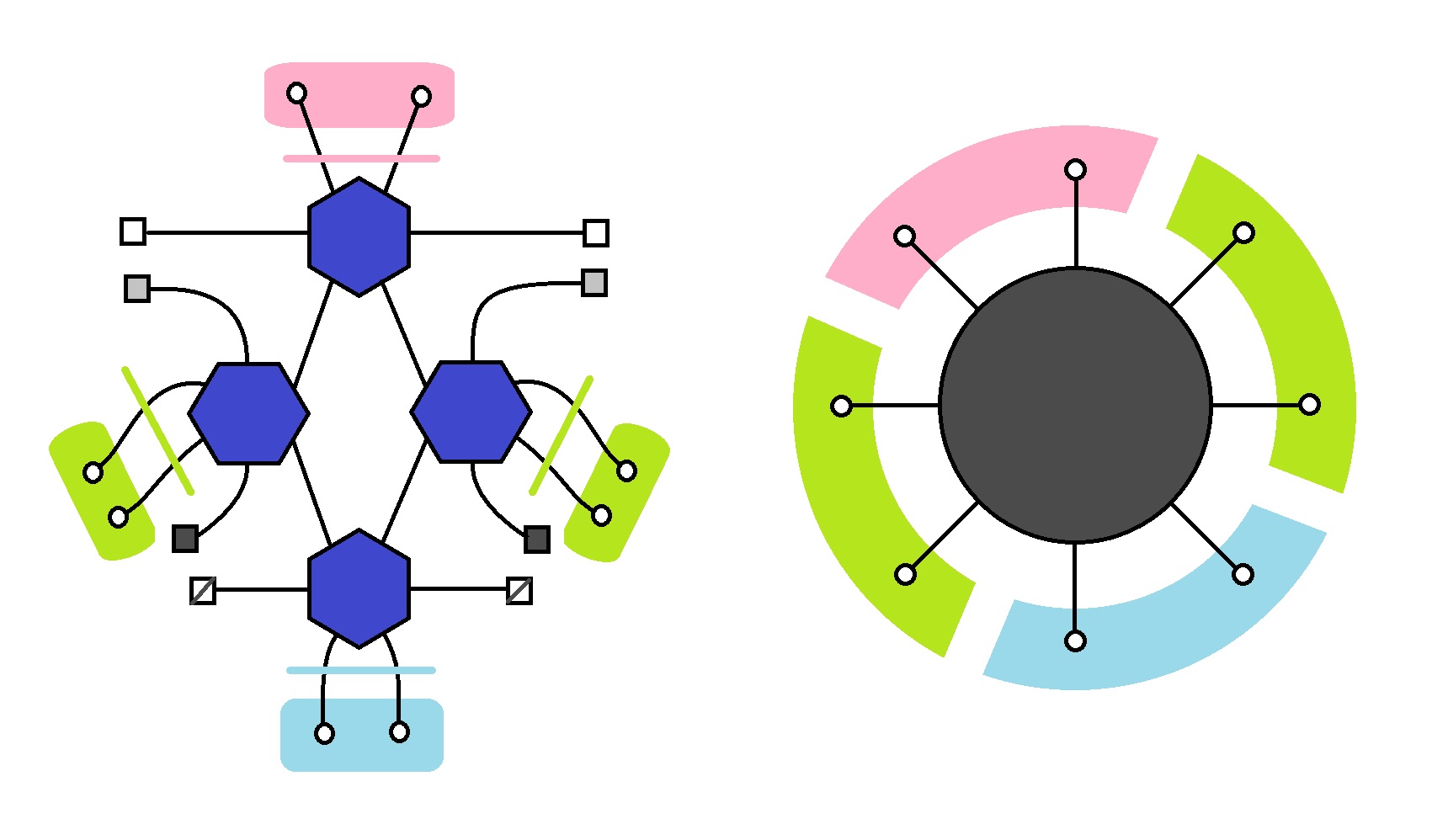}
\caption{(Left) The tripartite residual region of a network representing a very low temperature three-boundary wormhole. The coloured region containing circular boundary nodes depict the three different boundaries, and the cuts of the same colour indicate the horizon corresponding to each boundary. We compare the entanglement structures of the tensor network state and a state in the same Hilbert space defined by a Haar random state (right). }
\label{blob}
\end{figure}

The main interest in studying the tripartite residual region in these tensor network models, however, is that we can study the structure for small values of the discrete moduli, where the network in the causal shadow region is of modest size, and we can hope to analyse the resulting state on the legs crossing the horizon, and directly address questions about the nature of the entanglement. An example is shown in figure \ref{blob}. However, even for the smallest values of the moduli we are in a regime where there is no full classification of multipartite entanglement structures, so there is a shortage of general expectations to compare to. In \cite{Balasubramanian:2014hda}, the entanglement entropies obtained from holographic calculations were found to be consistent with those expected for random states in a reduced Hilbert space. This motivates us to compare the results obtained for the network in the causal shadow region to those for a Haar-random state on the legs crossing the minimal cuts, drawn as a blob in the right panel of figure \ref{blob}. 

We characterise the states by considering the entanglement entropy associated to each of the asymptotic regions, and by considering the logarithmic negativity for pairs of regions \cite{Horodecki:1998, 2016arXiv161108007E}. The logarithmic negativity for a density matrix $\rho$ on a Hilbert space $\mathcal{H}_{A}\otimes\mathcal{H}_{A^C}$ is defined as \cite{Plenio:2005cwa}
\begin{equation}
\label{logneg}
L = \log (2\mathcal{N} + 1)
\end{equation}
where $\mathcal{N}$ is the entanglement negativity defined to be,
\begin{equation}
\label{neg}
\mathcal{N}_A= \frac{ \| \rho^{T_A} \| - 1}{2}
\end{equation}
and where $\rho^{T_A}$ is the partial transpose of the density matrix $\rho$ on the factor $A$. If $\rho$ has components $\rho_{a b \; a' b'}$, The components of the partially transposed density matrix are $(\rho^{T_A})_{ab \; a'b'} = (\rho^{T_A})_{a'b \; ab'} $. The logarithmic negativity \eqref{logneg} provides an upper bound for the distillable entanglement, or the number of Bell pairs that can be distilled between the factor $A$ and its complement \cite{Vidal:2002cme}. Consider a Hilbert space of dimension $\chi^N$ (such as the state defined by an N-legged tensor with bond-dimension $\chi$), the maximum number of Bell pairs that could be ideally distilled from this state is $N_{\text{Bell}} = N \log_2 (\chi)$. When we have $L < N_{\text{Bell}}$ it indicates that there can be some component of the entanglement between $A$ and $A^C$ which is intrinsically multipartite. 

\begin{figure}
\centering
\includegraphics[width=\textwidth]{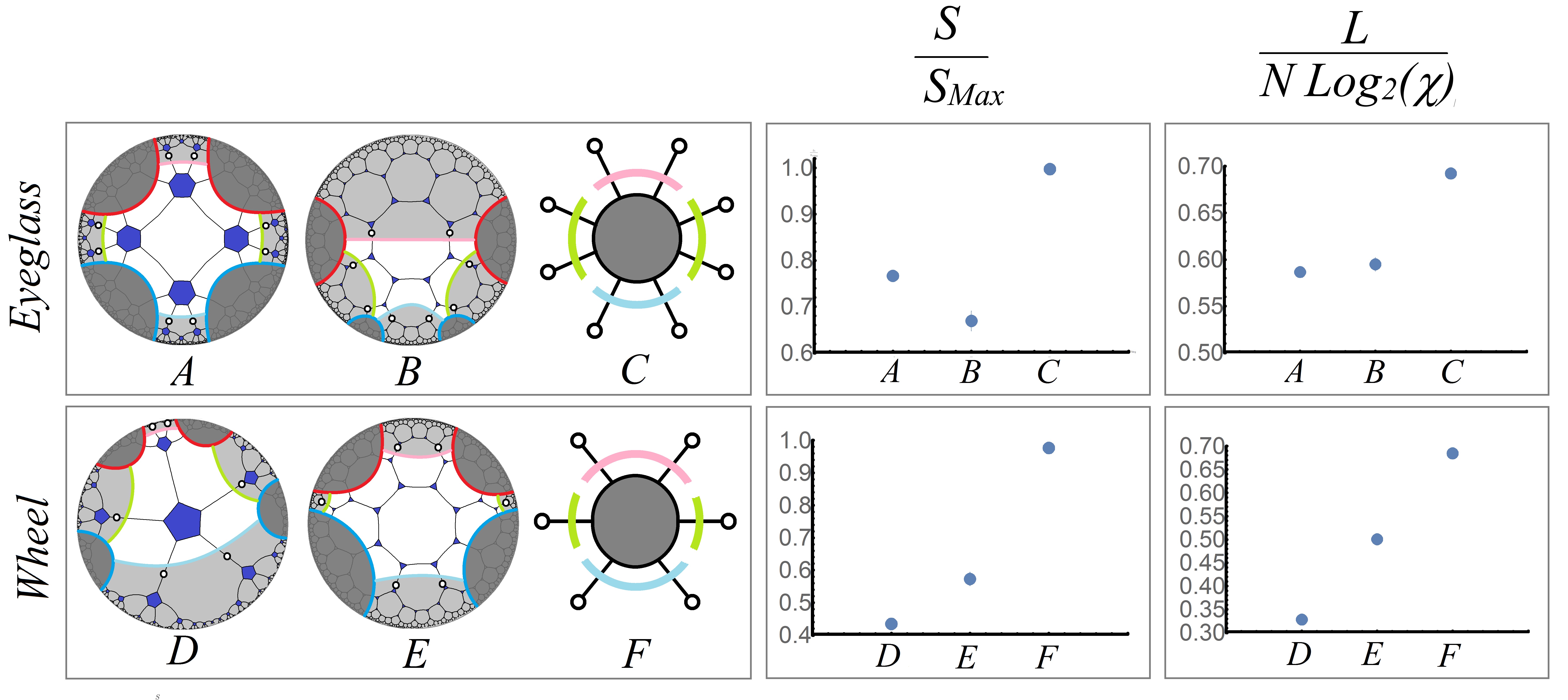}
\caption{Results comparing the entanglement structures of tripartite residual regions (highlighted regions of each network) of very low-temperature networks built out of Haar random tensors. We compare a pair of networks in each of the eyeglass (A and B) and wheel (D and E) regimes with a corresponding random states (C and F respectively) with the same dimensions. We compare the resulting entropy and logarithmic negativity on the light-blue factor in each case and compare the result with an appropriate random state, as shown in the plots on the right; error bars are barely perceptible. In the cases illustrated, the entanglement structure is quantitatively distinct to a random state. Here we take $\chi=3$. }
\label{tnvsblobs}
\end{figure}

\begin{figure}
\centering
\includegraphics[width=0.6\textwidth]{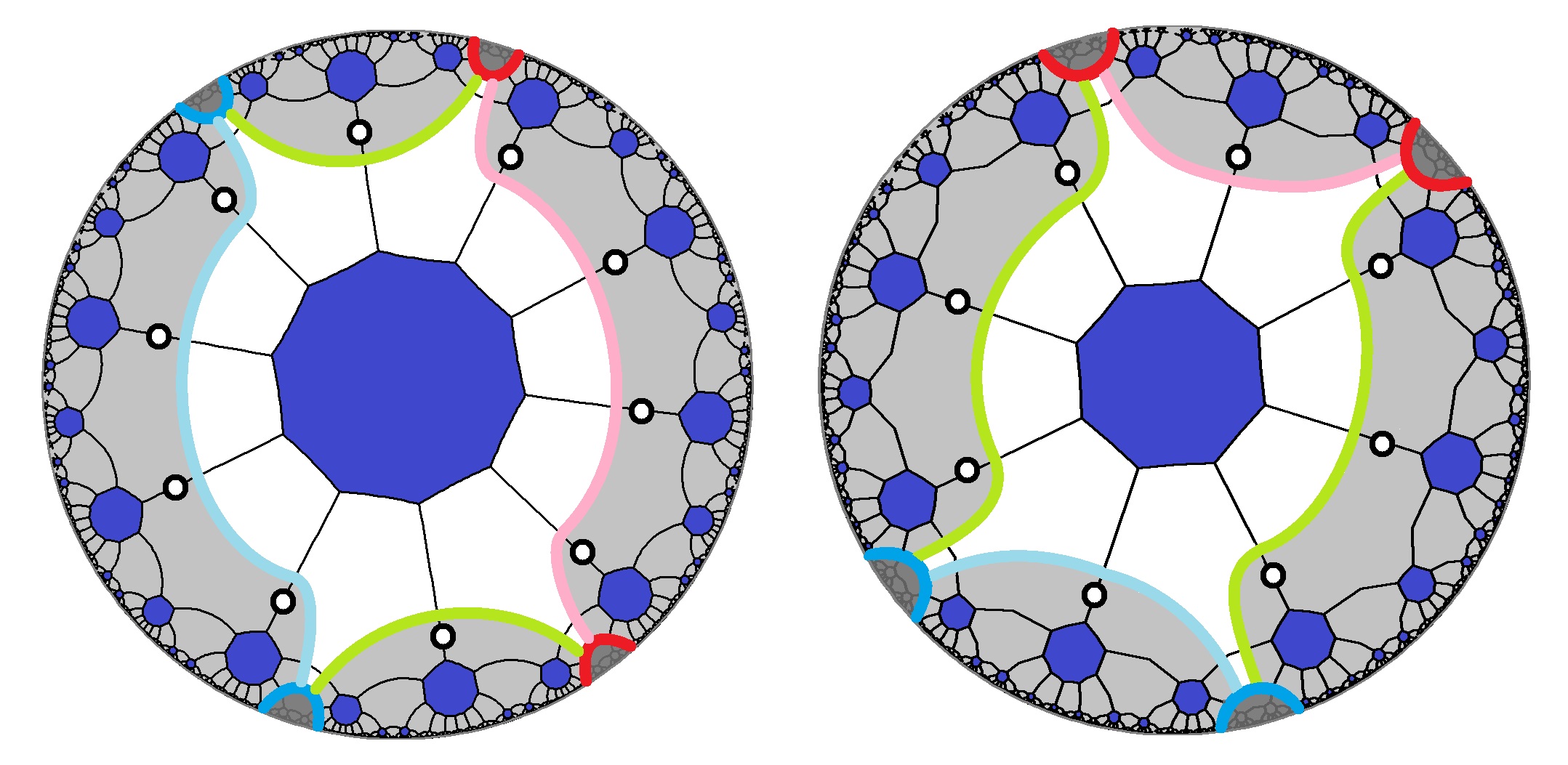}
\caption{Networks for low-T three-boundary wormholes for which the causal shadow (unshaded) is presicely a Haar random state, in the eyeglass (left) and wheel (right) regimes.}
\label{tnisblob}
\end{figure}

Results comparing the entanglement structure of tripartite residual regions for low-temperature wormholes built from a selection of tilings are shown in figure \ref{tnvsblobs}. We find that the distinction between these networks and random blobs depends strongly on the tiling. Not surprisingly, for choices of tilings where the causal shadow contains a single tensor, the results are as for a Haar random state, as is the case for the examples shown in figure \ref{tnisblob}. In contrast, networks for which the causal shadow region contains multiple tensors generically exhibit distinct entanglement structure, as with the networks shown in figure \ref{tnvsblobs}. The right-hand plots of figure \ref{tnvsblobs} correspond to non-vanishing logarithmic negativities on a single boundary factor. This is qualitatively unlike the $\text{GHZ}$ state, for which the logarithmic negativity on any single factor vanishes, and suggests the presence of some bipartite entanglement even in the state in this tripartite residual region. The fact that the logarithmic negativities do not reach their maximum value however implies a degree of intrinsically multipartite entanglement, as we expect. Though our results correspond only to relatively low bond-dimension, we expect the results for high bond-dimension should be at least qualitatively similar insofar as the entanglement structure of tensor network states is primarily dependent on the set of contractions within it.

\begin{figure}
\centering
\includegraphics[width=0.6\textwidth]{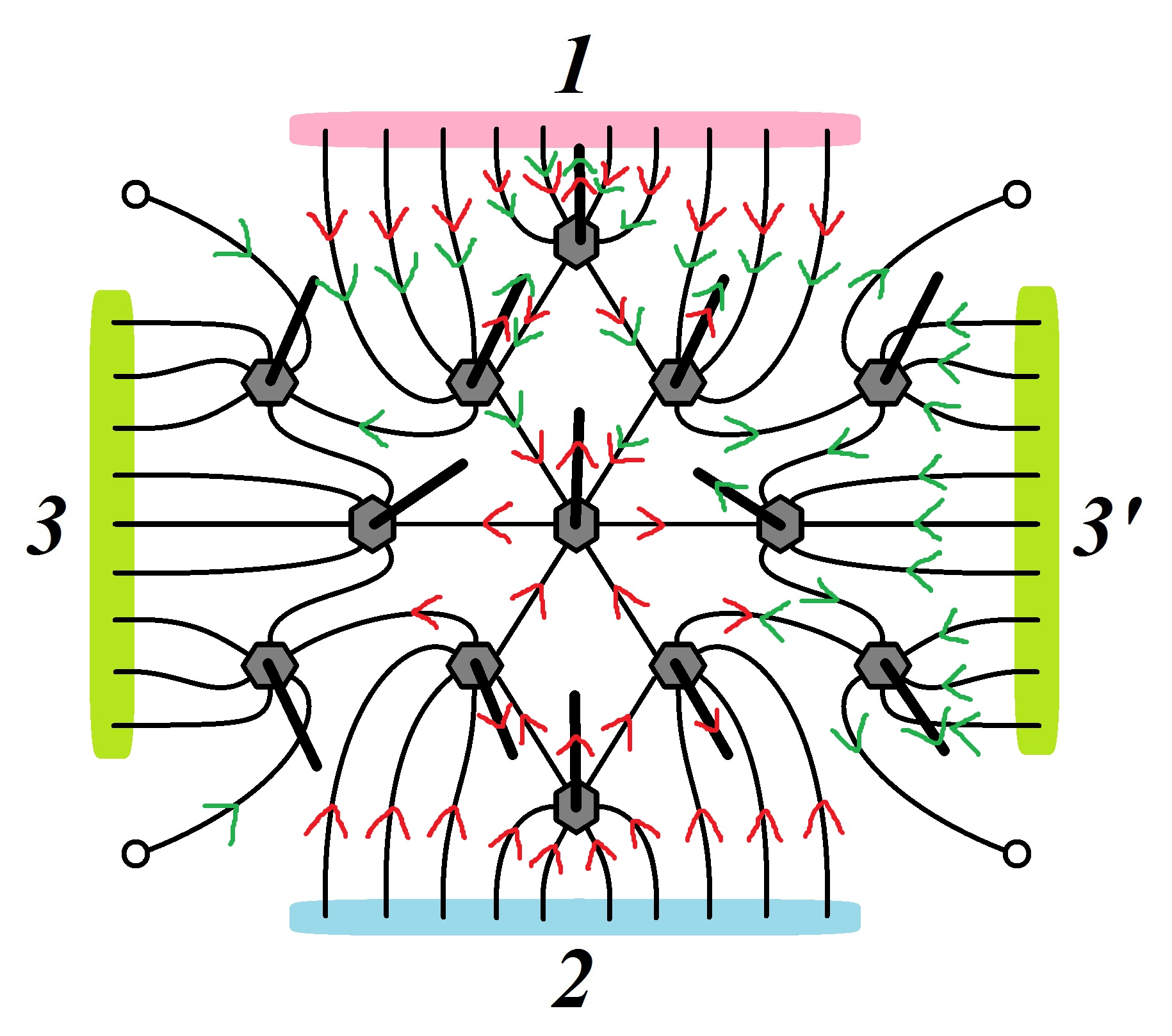}
\caption{A holographic septagon code for three-boundary wormhole in the wheel regime. Dangling bulk legs are indicated by thicker lines. The red and green arrow assignments depict our attempts to reconstruct the operator bulk site at the centre. The red arrow assignment corresponds to choosing the 1 and 2 for which we can reconstruct the central site. The green arrow assignment attempts to reconstruct the central site from just boundary 1 and 3' (half of boundary 3) and fails. The code thus behaves like a quantum secret sharing scheme, for which access to information, here pertaining to local bulk information, is only accessible with a sufficiently large share in hand; in this case being a sufficiently large portion of the boundary.}
\label{secret}
\end{figure}

It is also interesting to take a holographic code network, and consider the reconstruction of operators in the causal shadow region in terms of boundary operators. This requires us to push the operator back to the boundary through the tensors in the network. If there are self-contractions on the tensors this can lead to obstructions, but in general we find results that are consistent with the expectation that we can reconstruct an operator in this region from a subset of the boundary including more than half the legs along the boundary of the causal shadow region. An example is given in figure \ref{secret}. 

\begin{figure}
\centering
\includegraphics[width=0.8\textwidth]{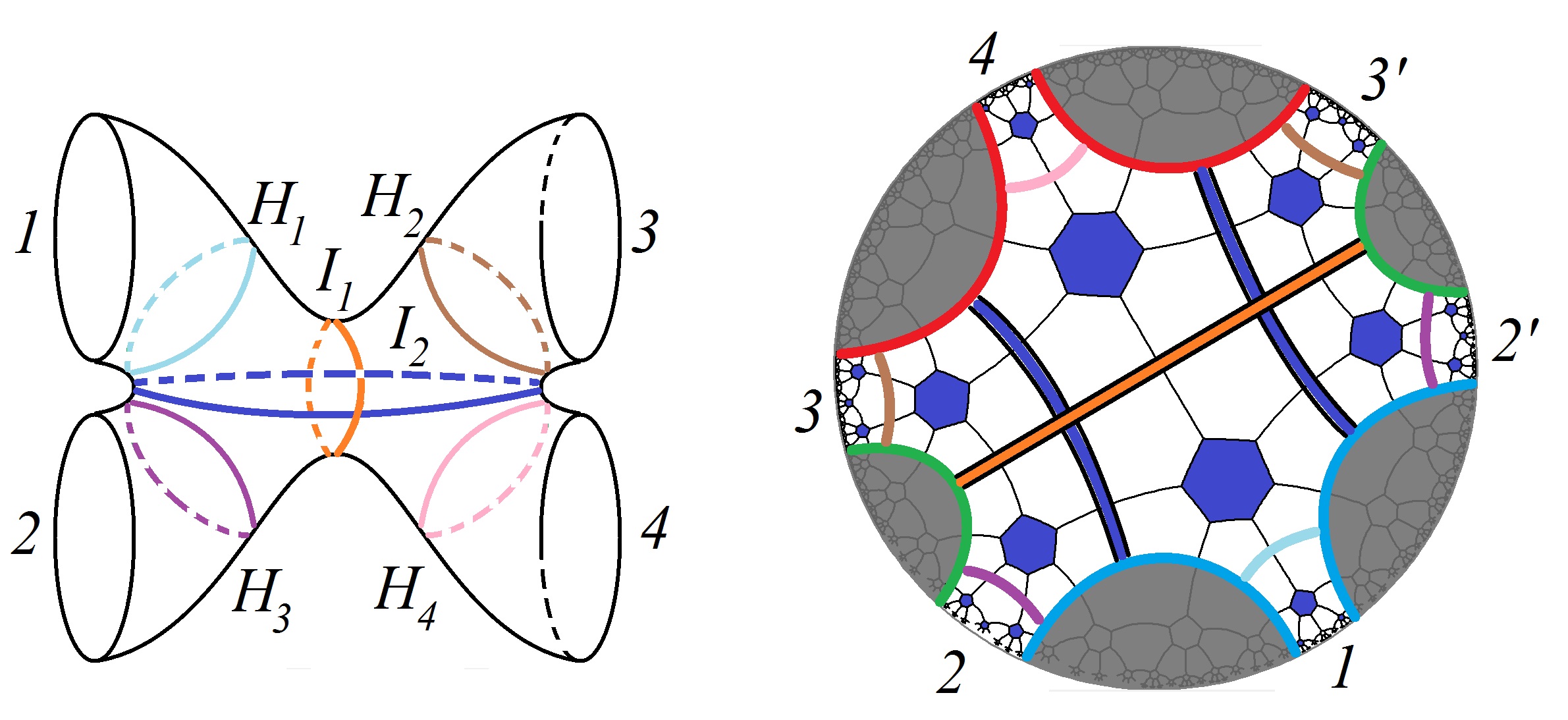}
\caption{A four boundary wormhole with horizons $H_a$, $a=1,...4$ (cyan, brown, pink, purple) and internal cycles $I_b$, $b=1,2$ (orange,blue) depicted.  Here, $I_1 < H_1 + H_2, H_3 + H_4$ and so $I_1$ is the RT surface for regions 1 \& 2 and regions 3 \& 4. A corresponding holographic state (right) is depicted, with colour-coded horizons and emphasised internal cycles indicated. Having $|I_1| < |H_1| + |H_3|$ constrains $S(1 \cup 2) \leq |I_1| \ln\chi = 3 \ln\chi$. In contrast, for a Haar random blob we would have $S(1 \cup 2) \leq \ln(dim( \mathcal{H}_1 \otimes \mathcal{H}_2 )) = 5 \ln\chi$.}
\label{fbs}
\end{figure}

We can also consider cases with more boundaries. These will then contain internal cycles, which can lead to important differences from the Haar random blob structure for the multipartite residual region. Short internal cycles constrain the maximum value of the entropy associated to a given subregion to \emph{less} than the typical maximal value. One example is the four-boundary wormhole with a short internal cycle. In the continuous case it is easy to see that sufficiently short internal cycles are dominant RT surfaces, as illustrated in figure \ref{fbs} (left). In a corresponding network this amounts to a constraint on the maximum entropy of the state on a set of boundaries for which the internal cycle is shorter than it's corresponding greedy geodesic, as illustrated in figure \ref{fbs} (right). 

\begin{figure}
\centering
\includegraphics[width=0.7\textwidth]{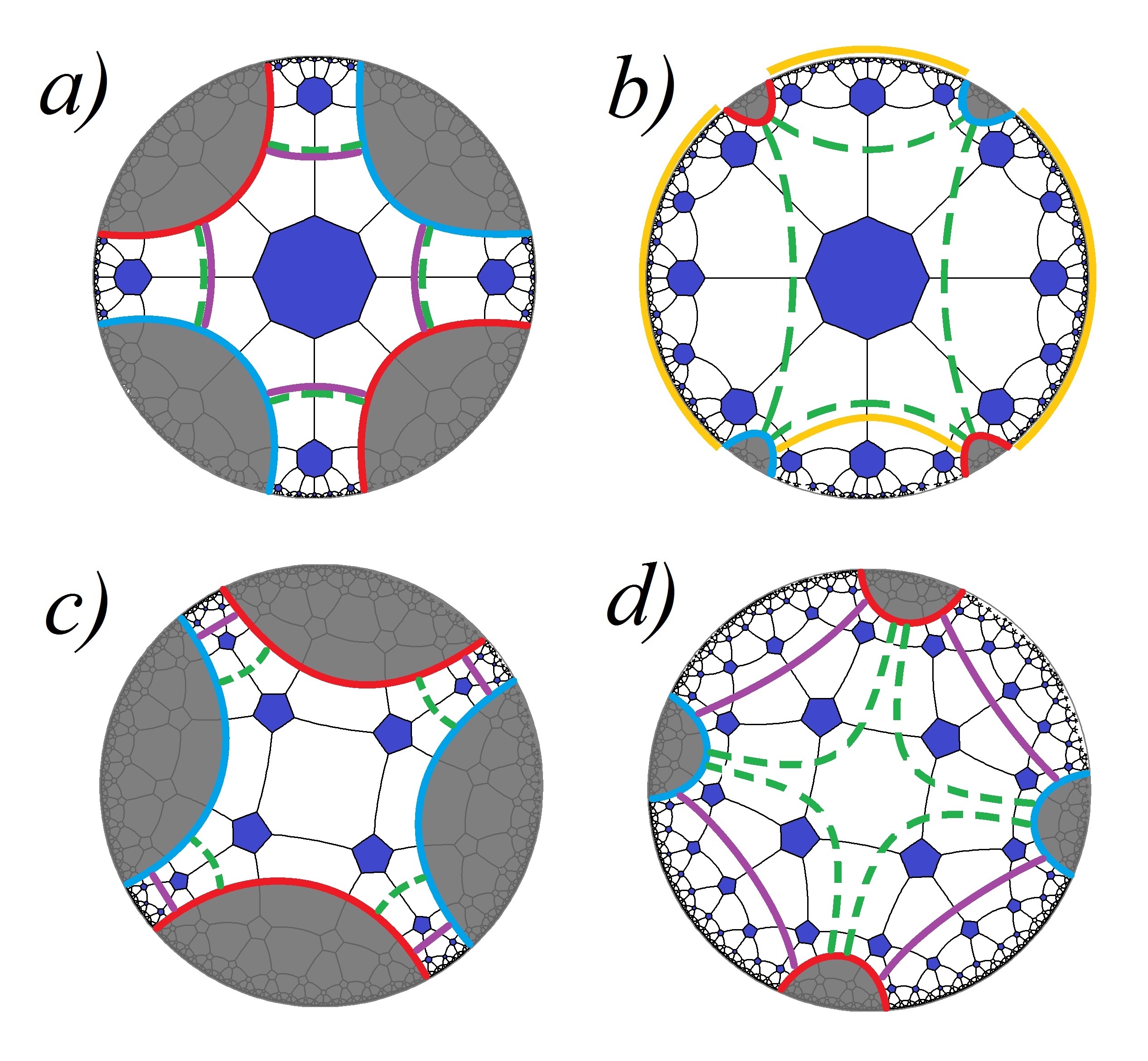}
\caption{low-T torus wormhole networks. a) A very low-T octagon state for which the minimal cut (dashed, green) wrapping the throat coincides with the greedy geodesic for the whole boundary (purple). Here we can't reconstruct the central tensor encoding the causal shadow. b) A higher-T octagon state for which the greedy geodesic is trivial meaning we can construct the central tensor lying behind the minimal cut from the whole boundary as we expect. In fact, we can reconstruct the tensor at the centre from only a subset of the boundary, as depicted by the yellow greedy geodesic associated to the portion of the boundary highlighted in the same colour, which passes through the central tensor. This mimics the anticipated high-T behaviour. c) A very low-T pentagon state with tiling artifacts in which the residual region does not reach the minimal cut. d) A higher-T pentagon state with the same tiling artifacts wherein we cannot reconstruct the region behind the minimal cut even with the whole boundary, because its corresponding greedy geodesic cannot penetrate the identifcation.}
\label{tori}
\end{figure}

Another example with internal topology is the torus wormhole. We find that networks constructed on tilings of the torus wormhole exhibit the expected features of their continuum analogues, up to tiling artefacts like those shown in figure \ref{dbtz}. In particular, we expect the torus networks to have a minimal cut homologous to the boundary that wraps the throat of the torus. Behind this region is the causal shadow with the topology of a torus. We expect that for low-temperature torus wormholes, the causal shadow region is reconstructible only with the entire boundary, whereas for higher temperatures we expect to be able to reconstruct information in the causal shadow with only subsets of the boundary. Examples of this behaviour are illustrated in figure \ref{tori} a) and b). 

\section*{Acknowledgements} 
We thank Danny Vagnozzi and Sichen Li for collaboration on initial stages of this project. AP is supported by an STFC studentship. SFR is supported by STFC under grant number ST/L000407/1. The illustrations of tilings were prepared using the Hyperbolic Tesselations package from http://dmitrybrant.com/2007/01/24/hyperbolic-tessellations. 

\bibliographystyle{JHEP}
\bibliography{multiboundary}

\end{document}